\documentclass[amsmath,amssymb,11pt]{article}
\usepackage{jheppub2}
\pdfoutput=1
\usepackage{graphicx,epsfig}
\usepackage{float}
\usepackage{subfig}
\usepackage{subfloat}
\usepackage[normalem]{ulem}

\usepackage[utf8]{inputenc}
\usepackage{hyperref}
\usepackage{caption}
\usepackage{color}
\usepackage{overpic}
\usepackage[dvipsnames]{xcolor}

\newcommand*{\affmark}[1][*]{\textsuperscript{#1}}

\definecolor{pink1}{rgb}{0.858, 0.188, 0.478}

\newcommand{\vol}{\mathcal{V}}

\def\beq{\begin{equation}}
\def\eeq{\end{equation}}

\title{Holographic CFT Phase Transitions  and  Criticality for Rotating AdS Black Holes}
\author{Moaathe Belhaj Ahmed,\affmark[1]}
\author{Wan Cong,\affmark[2]}
\emailAdd{wan.cong@univie.ac.at}
\author{David Kubiz\v{n}\'{a}k,\affmark[3,4]}
\emailAdd{david.kubiznak@matfyz.cuni.cz}
\author{Robert B. Mann,\affmark[1,4]}
\emailAdd{rbmann@uwaterloo.ca}
\author{\\and Manus R. Visser\affmark[5]}
\emailAdd{mv551@cam.ac.uk}

\affiliation{\affmark[1]Department of Physics and Astronomy, University of Waterloo,\\
 Waterloo, Ontario, N2L 3G1, Canada\\  
 \affmark[2]Faculty of Physics, University of Vienna, Vienna, Austria\\ 
\affmark[3]Institute of Theoretical Physics, Faculty of Mathematics and Physics,
Charles University, \\ V Hole\v{s}ovi\v{c}k\'{a}ch 2, 180 00 Prague 8, Czech Republic\\  
\affmark[4]{Perimeter Institute for Theoretical Physics, 31 Caroline St., Waterloo, Ontario, N2L 2Y5, Canada}\\
\affmark[5]Department of Applied Mathematics and Theoretical Physics,\\ University of Cambridge,
Wilberforce Road, Cambridge CB3 0WA, United Kingdom\\}

\abstract{ 
 Employing the novel exact dictionary between the   laws of extended  black hole thermodynamics and the laws of the dual CFT, 
we study the    extended thermodynamics   for     CFT states   that are dual to neutral singly-spinning asymptotically AdS black holes  in $d$  bulk spacetime dimensions. On the field theory side we include two   independent pairs of thermodynamic conjugate variables: the  central charge-chemical potential term and the   pressure-volume term. %, both of which can be varied independently. 
In this setting   
we uncover various phase transitions and critical behaviour in the CFT, focusing on three different thermodynamic ensembles.  Namely, for fixed angular momentum and central charge, we show there is a Van der Waals-like criticality for $d=4,5$ and reentrant phase transitions for $d\ge 6$. At fixed angular velocity  and central charge, there is a first-order (de)confinement phase transition in all dimensions $d \ge 3$.
Finally, at fixed angular momentum and chemical potential we find a plethora of  zero-order phase transitions and unstable phases in both $d=4$ and $d=6$.
}
\begin{document}

\maketitle

\section{Introduction}

One of the main advantages of  holographic duality is that puzzling features of black holes can be studied in the dual field theory, and vice versa. The best understood example of such a duality is the AdS/CFT correspondence \cite{Maldacena:1997re,Witten:1998qj,Gubser:1998bc}, where Anti-de Sitter (AdS) black holes have been argued to be equivalent to thermal states in the dual conformal field theory (CFT). This correspondence can be used as an argument for the unitarity of the evaporation of a black hole, since the dual CFT is a standard unitary gauge theory, albeit with a large number of color degrees of freedom $N$.  
%One of the holographic dictionaries 
 The holographic dictionary
states that the thermodynamics of AdS black holes is completely equivalent to the thermodynamics of the dual CFT. For instance,  the entropy and temperature of   a black hole match   with the thermal entropy and temperature of the dual CFT, respectively. Moreover, the  Hawking--Page first-order  phase transition \cite{Hawking:1982dh} between a large black hole and thermal AdS spacetime   corresponds to  the confinement/deconfinement phase transition of a quark gluon plasma~\cite{Witten:1998zw}.

 In recent years, the thermodynamics of AdS black holes has been shown to feature  a rich range of phenomena, including Van der Waals   type phase transitions  for charged AdS black holes  \cite{Chamblin:1999tk, Chamblin:1999hg, Cvetic:1999ne,Kubiznak:2012wp}, polymer transitions~\cite{Dolan:2014vba}, reentrant phase transitions
\cite{Altamirano:2013ane,Frassino:2014pha}, triple points
\cite{Altamirano:2013uqa,Wei:2014hba},  superfluid transitions \cite{Hennigar:2016xwd}, and most recently multicriticality \cite{Tavakoli:2022kmo,Wu:2022bdk}. 
These phenomena %\tcr{\sout{arise because of the extension of the thermodynamic phase space to include}}
 have been discovered in the context of extended phase space thermodynamics where the (negative) cosmological constant $\Lambda$ is treated as a dynamical variable, and identified with a (positive) thermodynamic pressure according to 
\cite{Kastor:2009wy,Dolan:2011xt,Dolan:2010ha,Cvetic:2010jb,Kubiznak:2014zwa} 
\be\label{P}
P=-\frac{\Lambda}{8\pi G_N}\,, \qquad  \Lambda=-\frac{(d-1)(d-2)}{2 L^2}\,, 
\ee
where   $L$ stands for the  AdS curvature radius, $G_N$ is the (fixed) Newton's constant, and $d$ denotes the number of bulk spacetime dimensions. This identification gives rise to a new pair of conjugate variables in the first law of bulk thermodynamics -- a   pressure-volume term, $+V\delta P$, arises. The corresponding theory has come to be called \emph{extended black hole thermodynamics}, or \emph{black hole chemistry} (see \cite{Kubiznak:2016qmn} for a review).
In particular, for a charged and multiply-spinning AdS black hole 
the    first law    and  the   generalised Smarr relation take the following form, respectively: 
\ba
\delta M&=&T_H \delta S+\Phi \delta Q 
+\sum_i \Omega_i \delta J_i
+ V \delta P\,,  \label{flaw}\\
M&=&\frac{d-2}{d-3} \bigl( 
%\frac{\kappa A}{8 \pi G_N} 
TS+ \sum_i\Omega_i J_i \bigr)+\Phi Q-\frac{2}{d-3}PV\,. \label{smarr}
\ea  
Here, $M$ stands for the mass of the black hole, $T_H$ for the Hawking temperature, $S$ for the Bekenstein-Hawking entropy,  $Q$ the  electric charge and $\Phi$  the conjugate  electrostatic potential. The angular momenta of the black hole are denoted by $J_i$, and their respective conjugate quantities are the  relative angular velocities between horizon and infinity~$\Omega_i$ \cite{Gibbons:2004ai}, and 
$V$ is the black hole thermodynamic volume -- a quantity conjugate to the pressure~$P$.
%\be\label{Vdef}
%V=\Bigl(\frac{\partial M}{\partial P}\Bigr)_{S,Q,J_i}\,.
%\ee

Very recently \cite{Ahmed:2023snm}, a precise match has been found between the laws of extended black hole themodynamics and the laws governing the dual CFT.
The key idea for this identification is to respect the conformal symmetry of the dual CFT, treating the 
AdS boundary conformal factor $\omega$ as a new thermodynamic parameter, so that the CFT volume ${\cal V}$ and the CFT central charge $C$ can be varied independently, without  needing to introduce  
%introducing 
a variable Newton's  constant, as done in \cite{Karch:2015rpa, Visser:2021eqk, Cong:2021fnf, Zeyuan:2021uol, Wang:2021cmz, Lobo:2022eyr}. Namely, in this framework the conformal completion of the bulk AdS spacetime reads as  
\be \label{CFTmetric1}
ds^2= \omega^2\Bigl(-dt^2+L^2d\Omega_{d-2}^2\Bigr)\,,
\ee
where   $\omega$ is an `arbitrary'  dimensionless  conformal factor that is free to vary, reflecting the conformal symmetry of the boundary theory. %, and will be treated as a thermodynamic parameter. 
Focusing on the spherical case, $d \Omega_{d-2}^2 \!$ is the metric on a unit ($d-2$)-dimensional sphere, with the corresponding volume $\Omega_{d-2}$. We take $\omega$ to be independent of the boundary coordinates, in which case the CFT volume   reads
\be \label{volume}
{\cal V}=\Omega_{d-2}R^{d-2}\,,
\ee 
where $R=\omega L$ is the variable curvature radius of the manifold where the CFT lives. The variation of the CFT volume ${\cal V}$ is then obviously independent of the variation of the central charge $C$, which for   Einstein gravity is dual to  
\be\label{C}
C=\frac{\Omega_{d-2}L^{d-2}}{16\pi G_N}\,,
\ee
even when   Newton's constant $G_N$ is held fixed. 
Employing the following AdS/CFT dictionary:
\ba \label{extendeddictionary}
 % S&=& \frac{A}{4 G_N}\,,\quad 
  E=\frac{M}{\omega}\,,\quad  T=\frac{T_H}{\omega}\,,\quad \tilde\Omega=\frac{\Omega}{\omega}\,,  \quad \tilde \Phi=\frac{\Phi\sqrt{G_N}}{\omega L}\,,\quad \tilde Q=\frac{QL}{\sqrt{G_N}}\,,
\ea
it is easy to show  that the bulk first law \eqref{flaw} is dual to \cite{Visser:2021eqk,Ahmed:2023snm}:
\be
    E= T\delta S+\tilde \Omega \delta J+\tilde \Phi \delta \tilde Q+\mu \delta C-p\delta {\cal V}\,,\label{j1}
\ee
accompanied by the following two relations for the chemical potential $\mu$ associated to the central charge and for the pressure $p$, respectively, 
\ba
\mu&=&\frac{1}{C}(   E -    T S-\tilde \Omega J-\tilde \Phi \tilde Q)\,, \label{j2}\\
p&=& \frac{  E}{(d-2){\cal V}}\,,\label{j3}
\ea 
known as the    Euler relation  and the equation of state for   CFTs, respectively. 
This Euler equation holds for any large-$N$ gauge theory, and  differs from the standard one in thermodynamics 
 in that it does not contain a $p \mathcal V$ term. 
 In the high-temperature
or large-volume regime, i.e. $R  T\gg 1$, the $\mu C$ term
becomes equal to $- p{\cal V}$, and  \eqref{j2}  becomes the standard thermodynamic Euler relation \cite{Visser:2021eqk}. In Ref. \cite{Cong:2021jgb} we provided an extensive study of the   extended  thermodynamics of CFT states dual to charged, nonrotating AdS black holes.

 It is the purpose of this paper to explore the implications of this proposal for rotating thermal CFT states  that are dual to uncharged, singly-spinning AdS black holes in the bulk.
In particular, we shall focus on the following three ensembles that feature interesting phase behavior: 
%\tcr{\sout{following recent work in which the holographic dual behaviour of charged AdS black holes was studied \cite{Cong:2021jgb}.}}  We shall \tcr{focus on}
%singly-rotating uncharged AdS black holes in $d$ dimensions so as to better clarify the role of rotation in the duality. 
%As in the charged case \cite{Cong:2021jgb}, we shall explore the various thermodynamic ensembles possible in this scenario.  Amongst these, we find that only the following three ensembles admit interesting phase behaviour:
    \begin{equation}
     \begin{aligned}  
    &\text{fixed} \quad  (J, {\cal V}, C): \qquad &&F \equiv E -   T S   \,,\\
      &\text{fixed} \quad  (\tilde \Omega, {\cal V},C)\,: \qquad &&W \equiv E-   T S - \tilde \Omega J  \,,\\
       &\text{fixed} \quad  (J, {\cal V}, \mu)\,: \qquad &&G \equiv E -   T S-     \mu C \,,
     \end{aligned}
 \end{equation}
 where   $F$, $W$, and $G$ are the corresponding free energies of the respective ensembles. These ensembles are analogous to the three   ensembles studied in \cite{Cong:2021jgb} for thermal CFT states  dual to charged, nonrotating  AdS black holes, for which we found   interesting phase behaviour.
For the present rotating case, in the first (`canonical') ensemble, we shall show that there is a Van der Waals-like criticality for $d=4,5$, and reentrant phase transitions for $d\ge 6$. In the `grand canonical ensemble', at fixed angular velocity  and central charge, there is a first-order (de)confinement phase transition in all dimensions $d \ge 3$, following closely what happens in the bulk (see App.~\ref{appA}).
%\tcr{characterized, however, by a remarkably different from bulk phase diagram {\bf if this remains true}.    
Finally, the behavior of the free energy in the last ensemble, characterized by fixed angular momentum and chemical potential, is rather complex and unprecedented. It seems to indicate the presence of a plethora of zero-order phase transitions and unstable phases in both $d=4$ and $d=6$. However, one should be a bit cautious about the precise interpretation of these results, as this ensemble is novel and may come with presently unknown phases and instabilities that our analysis did not take into account.

Let us finally stress that the current analysis has a certain overlap with recent investigations \cite{Zeyuan:2021uol,Gao:2021xtt,Wang:2021cmz,Cong:2021jgb,Zhao:2022dgc,Kong:2022gwu,Gong:2023ywu}. In particular,  the first two ensembles have been studied in $d=4$~\cite{Gao:2021xtt}, in the context of the so-called {\em restricted phase space (RPS)} formalism. While similar in many technical aspects to our approach, the physical interpretation of RPS is very different from ours. Namely, in RPS the cosmological constant $\Lambda$ is fixed in the bulk, while the gravitational constant $G_N$ is varied.\footnote{
Let us note here that  a number of papers 
  have considered the possibility of adding 
  Newton's constant $G_N$  to  the  \emph{extended thermodynamic phase space}, e.g.  \cite{Kastor:2010gq,Kastor:2014dra,Karch:2015rpa,Sarkar:2020yjs,Visser:2021eqk,Cong:2021fnf}. Since the inclusion of quantum corrections seems to indicate that $G_N$ varies along the renormalization group flow, its variation in thermodynamics is perhaps plausible. However, such a treatment seems a bit problematic, in part since $G_N$ is a constant of nature and varying $G_N$ does not correspond to the original black hole chemistry. For this reason we try to avoid it in this paper.}
Moreover, since on the CFT side in RPS one fixes $\omega=1$, it implies that the CFT volume has to be held fixed and only the central charge remains a thermodynamic variable. Contrary to this, in our case, we hold   Newton's constant fixed in the bulk and only vary $\Lambda$. However, by introducing general $\omega$ on the CFT side, both the CFT volume and the central charge are independently varied, subject to the two restrictions \eqref{j2} and \eqref{j3}.

Our plan for the remainder of the paper is as follows. In Sec.~\ref{sec2} we review  singly-spinning AdS black holes in all dimensions and discuss their respective bulk and boundary thermodynamic quantities. Sec.~\ref{sec3} contains our main results regarding the phase behavior of the three thermodynamic ensembles. Sec.~\ref{sec4} is devoted to   discussion of our results and conclusions. App.~\ref{appA} reviews the grand canonical ensemble behavior of the singly-spinning black holes in the bulk, which is also compared to the (markedly different) fixed electrostatic potential ensemble for charged nonrotating AdS black holes. Additional technical details regarding the study of the $(J,{\cal V}, \mu)$ ensemble are gathered in App.~\ref{appB}.

%\mv{\bf 
% Also mention the old bulk interpretation in terms of pressure $P$ and volume $V$, and the new mixed bulk/boundary interpretation in terms of $(P,V)$ and $(\mu,C)$, where the chemical potential is different from the one considered here.
%} \tcr{\bf I am not sure we really need to 
%do it here but do not object strongly...}

\section{Holographic thermodynamics of Kerr-AdS black holes}\label{sec2}

In this section we relate the extended thermodynamics of rotating black holes in AdS to the extended thermodynamics of the dual CFT. 
%\tcr{\sout{ following  recent work for the charged case \cite{Visser:2021eqk,Cong:2021fnf,Cong:2021jgb}.}} 
We keep the number of (bulk) spacetime dimensions $d$ arbitrary in this section, while the detailed analysis of the CFT phase behavior in the next section will be carried out in $d=4$ and $d=6$, i.e. AdS$_4$/CFT$_3$ and AdS$_6$/CFT$_5$. 

 \subsection{Extended bulk thermodynamics}
 
We consider neutral  singly-spinning black holes in asymptotically AdS spacetime, also known as Kerr-AdS black holes. These form a two parameter family of solutions to the vacuum Einstein equations with a negative cosmological constant, which were constructed   in four dimensions by Carter  \cite{Carter:1968ks} and later generalized to higher dimensions in \cite{Hawking:1998kw,Gibbons:2004uw,Gibbons:2004js}. %Kerr-AdS is a solution to general relativity with a negative cosmological constant $\Lambda$
 %\begin{equation}
 %	I = \frac{1}{16\pi G} \int d^{d+1 } x\sqrt{-g} \left (  R - 2 \Lambda    \right), \qquad \Lambda = - \frac{d (d-1)}{ 2L^2},.
% \end{equation} 
In Boyer-Lindquist coordinates, the  Kerr-AdS line element in $d$ spacetime dimensions reads 
 \begin{equation}\label{metric}
 	ds^2=-  \frac{\Delta }{\rho^2} (dt - \frac{a}{\Xi} \sin^2 \theta d \phi)^2 + \frac{\rho^2}{\Delta} dr^2 + \frac{\rho^2}{\Sigma} d \theta^2 + \frac{\Sigma \sin^2 \theta}{\rho^2} \left (a dt - \frac{r^2 + a^2}{\Xi} d \phi \right)^2 + r^2 \cos^2 \theta d \Omega^2_{d-4}\,,
 \end{equation}
 where $d \Omega_{d-4}^2$ is the metric on the round unit $d - 4$ sphere, and the   various metric functions   are given by
 \begin{equation}
 	 \Delta = (r^2 + a^2)\left (1 + \frac{r^2}{L^2} \right) - \frac{2 m}{r^{d-5}} , \quad \Sigma = 1 - \frac{a^2}{L^2} \cos^2 \theta  , \quad \Xi = 1 - \frac{a^2}{L^2}  , \quad \rho^2 = r^2 + a^2 \cos^2 \theta.
 \end{equation}
Here $L$ is the AdS curvature radius given in \eqref{P}, $m$ is the mass parameter and $a$   the rotation parameter. The mass parameter $m$ can be expressed in terms of the other parameters and the  outer horizon radius $r_h$   (the largest  positive real   root of $\Delta =0$) as 
 \begin{equation}
 	m =\frac{r_h^{d-5}}{2} (r_h^2 + a^2 ) \left ( 1 + \frac{r_h^2}{L^2}\right) \,.
 \end{equation} 
 
 The expressions for the thermodynamic variables of the Kerr-AdS black hole   in terms of the variables $(r_h,a,L)$ are well known in the literature (see e.g. \cite{Gibbons:2004uw}). Here we provide   a quick summary. The mass and angular momentum of   Kerr-AdS black holes are given by
 \begin{equation}
 \label{mass}
 	M=  \frac{\Omega_{d-2}}{4 \pi G_N} \frac{m}{\Xi^2} \left ( 1 + \frac{(d-4)\Xi}{2} \right)\,, \qquad  	J= \frac{\Omega_{d-2}}{4 \pi G_N} \frac{ma}{\Xi^2}\,.
 \end{equation}
% Angular momentum
% \begin{equation}
% 	J= \frac{\Omega_{d-2}}{4 \pi G} \frac{ma}{\Xi^2}
% \end{equation}
The angular velocity of the black hole  horizon relative to spatial infinity is 
%,  relative to a non-rotating frame at infinity, is
 \begin{equation}
 	\Omega = \frac{a}{L^2} \frac{r_h^2 + L^2}{r_h^2 + a^2} \,.
 \end{equation}
The Hawking temperature is proportional to   the surface gravity $\kappa$ according to
 \begin{equation}
 	T_H = \frac{\kappa}{2\pi} =  \frac{1}{2\pi} \left [ r_h \left ( 1 + \frac{r_h^2}{L^2}\right)  \left ( \frac{1}{a^2 + r_h^2} + \frac{d-3}{2 r_h^2}\right) -\frac{1}{r_h}\right]\,,
 \end{equation}
and
 \begin{equation}
 \label{S}
 	S=\frac{A}{4G_N} = \frac{\Omega_{d-2}}{4 G} \frac{r_h^{d-4}(a^2 + r_h^2)}{\Xi}
 \end{equation} 
 is the Bekenstein--Hawking entropy, proportional to the area $A$ of the outer event horizon.

 The thermodynamic volume can either be calculated from the extended first law \eqref{flaw}, or from the Smarr relation \eqref{smarr}. In either case we recover \cite{Dolan:2011xt, Cvetic:2010jb} 
 \begin{equation}
 	V =  \frac{r_h A}{d-1} \left [ 1 + \frac{a^2}{\Xi} \frac{1 + r_h^2/L^2}{(d-2)r_h^2}\right]\,.
 \end{equation}
%The \textit{generalised Smarr formula} \eqref{smarr} becomes \cite{Visser:2021eqk}
% \begin{equation}
% \label{smarrKAds}
% 	M  =\frac{d-2}{d-3} \left( \frac{\kappa A}{8 \pi %G_N}  + \Omega J \right) - \frac{1}{d-3} %\frac{\Theta \Lambda}{4 \pi G_N}\,,
% \end{equation}
% where $\Theta$ is the quantity conjugate to the cosmological constant  $\Lambda$ in an extended first law (see below) and is equal to minus   the thermodynamic volume, $\Theta = - V$.
% By combining the expressions for the thermodynamic quantities in terms of the horizon radius, one can find the following expression 
Alternatively, this expression can also be 
%c
%which depends on the horizon coordinate radius $r_h$, can also be 
computed by using a geometric approach. Indeed, it can be defined either in terms of surface integrals of the Killing potential \cite{Kastor:2009wy} (see \cite{Cvetic:2010jb} for an explicit calculation using the proper gauge fixing of the Killing potential, as opposed to the background subtraction procedure proposed in  \cite{Kastor:2009wy}), or as    the (background subtracted) Killing volume  \cite{Jacobson:2018ahi}
 \begin{equation}
 	V =  \int_{\Sigma_{\text{bh}}} |\xi| dV - \int_{\Sigma_{\text{AdS}}} |\xi| dV\,,
 \end{equation}
 where $|\xi|=\sqrt{-\xi \cdot \xi}$ is the norm of the horizon generating Killing vector $\xi = \partial_t + \Omega \partial_\phi$.  Let us now relate these bulk thermodynamic quantities to the boundary thermodynamic quantities  of the dual CFT.
 
 % Finally, the extended form of the first law of Kerr-AdS black holes, including    variations of $\Lambda$ and $G_N$, is given by \cite{Visser:2021eqk}
% \begin{equation}
% \label{bulk law}
% 	d M = \frac{\kappa}{8\pi G_N} dA + \Omega dJ +\frac{\Theta}{8\pi G_N} d \Lambda - \left ( M   - \Omega  J  \right)\frac{dG_N}{G_N}\,.
% \end{equation}
%The final term follows from the fact that the mass and angular momentum are inversely propotional to Newton's constant: $M, J \sim 1/G_N$. Both $\Lambda$ and $G_N$ can be viewed as coupling constants that may vary in the space of gravitational theories. We do not give them a thermodynamic interpretation separately, but rather relate them to 
%the central charge,
%a dimensionless quantity that characterizes the number of field degrees of freedom in the dual CFT. Hence, at this stage  the variations of $\Lambda$ and $G$  are   helpful bookkeeping devices in finding the correct extended first law in the dual CFT. %  In the next section we match this bulk first law to an extended first law in the dual CFT. 

%\input{Section2b.tex}

\subsection{Extended boundary thermodynamics}

%\MV{work on this..}
In the AdS/CFT correspondence the  dual CFT  lives on the conformal boundary of the asymptotically AdS spacetime.  According to \cite{Gubser:1998bc,Witten:1998qj}, the CFT metric is identified with the boundary metric  of the dual asymptotically AdS spacetime up to a Weyl rescaling, $g_{\text{CFT}} = \text{lim}_{\rho  \to \infty}  \lambda^2 (x) g_{\text{AdS}}$, where $\rho$ is a radial coordinate and $\lambda(x)$ is a Weyl scale factor. Following \cite{Gibbons:2004ai,Gibbons:2005jd}, we take the boundary metric to be that of the Einstein static Universe (up to a constant Weyl factor $\omega$), i.e.  the standard product metric on $\mathbb R \times S^{d-2}$:
%Our CFT lives on a boundary that is a rotating  Einstein Universe   with  angular velocity $\Omega$. 
\begin{equation}
\label{CFTmetric2}
ds^2= \omega^2\Bigl(-dt^2+L^2d\Omega_{d-2}^2\Bigr)\,.
\end{equation}
To see how this arises from an asymptotic limit of the Kerr-AdS metric, we perform the following coordinate transformation
\cite{Cardoso:2006wa}
\be
\varphi=\phi-\frac{a}{L^2}t\,,\quad 
\rho \cos\Theta=r \cos\theta\,,\quad \rho^2=\frac{1}{\Xi}(r^2\Sigma+a^2\sin^2\!\theta)\,,
\ee
where  we focus for simplicity on $d=4$ dimensions. This
brings the $m=0$ metric \eqref{metric} to the following form: 
\be
ds^2=-\Bigl(1+\frac{\rho^2}{L^2}\Bigr)dt^2+\frac{d\rho^2}{1+\frac{\rho^2}{L^2}}+\rho^2(d\Theta^2+\sin^2\!\Theta d\varphi^2)\,. 
\ee
By taking the limit $\rho \to \infty$ and multiplying with the Weyl factor   $\lambda =   \omega L/\rho$, we arrive at   the CFT metric   \eqref{CFTmetric2}.
Although the corresponding boundary metric is static, this is misleading as the regularity of the Euclidean  section of the Kerr-AdS black hole requires the  identification
\be
(t,\rho, \Theta, \varphi)\sim (t+i\beta, \rho, 
\Theta, \varphi+i\beta \Omega)\,, 
\ee
which gives rise to a rotating state on the boundary with linear velocity $v\equiv \Omega L$. Obviously, for 
\be\label{OmegaL}
\Omega>1/L 
\ee
the Einstein Universe on the boundary rotates faster than the speed of light, and the corresponding CFT seems ill defined.\footnote{It was 
shown in \cite{Hawking:1999dp}, that a weakly coupled CFT does not admit a well-defined partition function when $\Omega L>1$. While a good indication that `something' may go wrong for $\Omega>1/L$, our CFT is strongly coupled and the previous argument does not necessarily apply.
}
%We note that other interpretations for the boundary universe have been proposed \cite{Gibbons:2005jd}
%\tcr{\bf What is not clear to me at all is whether we have CFT in rotating Einstein Universe, or if we have "rotating state of CFT" in static Einstein Universe. I would naively expect that the two are different. However, from the bulk perspective I do not se a way to prefer one over the other, So which is it? In any case we need some comments on this!}  
In the bulk, this corresponds to the classical {\em superradiant instability} of the corresponding small black hole solutions, which have horizon radius $r_h<\sqrt{aL}$. As argued in \cite{Cardoso:2006wa}, the endpoint of such an instability corresponds to a `hairy black hole' with $\Omega L=1$. 

The existence of the bound \eqref{OmegaL} therefore imposes a restriction on the validity of the thermodynamic phase diagrams constructed below.     One possibility is to disregard the superradiant/faster than speed of light branches from the free energy diagrams completely (see black curves therein). As such states often minimize the free energy, doing so would completely modify the thermodynamic behavior of the system and would imply  novel phase diagrams (often characterized by additional zeroth-order phase transitions). However, physically it makes much more sense to assume that the superradiant/faster than speed of light branches will be replaced by the corresponding branches of stable `hairy' black holes/novel phases of CFT that are in some sense thermodynamically ``close to" the original $\Omega L >1$ branches. For this reason in what follows we `preserve' (apart from the grand canonical ensemble) the $\Omega L>1$ branches in the free energy diagrams and construct the corresponding phase diagrams as if these branches remained present. We expect that this gives a qualitatively better picture than disregarding these branches completely.

 The AdS formulae \eqref{mass}-\eqref{S}, together with the holographic dictionary    \eqref{C} and \eqref{extendeddictionary}, provide information about the (extended) thermodynamics of the dual large-$N$, strongly coupled CFT. In the next section, we will look at the implied thermodynamic phase behaviour of the CFT. For this purpose, it turns out to be convenient to introduce two dimensionless parameters,
   \begin{equation}
 	x \equiv \frac{r_h}{L}\,, \qquad \qquad z \equiv \frac{a}{L}\,,
 	\label{xz}
 \end{equation}
  with which we have
 \begin{equation}
 	\Sigma = 1 - z^2 \cos^2 \theta , \qquad \Xi = 1 - z^2, \qquad  
 	m L^{3-d} =\frac{x^{d-5}}{2}  (x^2 + z^2 ) \left ( 1 + x^2\right)\,.
 \end{equation} 
 In terms of $x$ and $z$, the CFT thermodynamic quantities are given by
 \begin{itemize}
     \item entropy:
 \begin{equation}
 \label{entropy}
 	S = 4 \pi C x^{d-4} \frac{x^2 + z^2}{1-z^2}\,,
 \end{equation}
  \item energy: 
 \begin{equation}
 	E =  \frac{1}{R} 4  C   \frac{ x^{d-5}   (x^2 + z^2 ) \left ( 1 + x^2\right)}{2(1 - z^2)^2} \left ( 1 + \frac{(d-4)(1-z^2)}{2} \right)\,,
 	\end{equation}
 	\item angular momentum:
 \begin{equation}
 \label{J}
 	J= 4  C   \frac{  z  }{(1-z^2)^2} \frac{x^{d-5}}{2}  (x^2 + z^2 ) \left ( 1 + x^2\right)\,,
 \end{equation} 	
 \item	temperature:
 	 \begin{equation}
 	 \label{T}
 	T =  \frac{1}{2\pi R} \left [ x \left ( 1 + x^2\right)  \left ( \frac{1}{x^2 + z^2} + \frac{d-3}{2 x^2}\right) -\frac{1}{x}\right] \,,
 \end{equation}
 \item angular velocity:
 \begin{equation}  \label{omega}
 	\tilde \Omega = \frac{z}{R } \frac{x^2 + 1}{x^2 + z^2} \,,
 \end{equation}
\item chemical potential:
 \begin{equation}
 \label{mu}
 	\mu = \frac{x^{d-5}\left(x^2-1\right)  \left(x^2+z^2\right)}{R
   \left(z^2-1\right)} \,.
   % \frac{x}{R   \left(z^2-1\right)^2} \left[\left(x^2-1\right) z^4+x^2- x^4 +\left(3 x^4+2 x^2-2 \pi  \left(x^2+1\right)^2+3\right) z^2\right ]
 \end{equation}
 \end{itemize}
%We can also express $y$ in terms of $x,T$ and $R$
%\begin{equation}
% y^2 = \frac{\tilde Q^2}{4 \alpha^2 (d-1)^2 C^2} =  x^{2d-4}+\frac{d}{d-2} x^{2d-2}-T x^{2d-3} \frac{4 \pi R}{d-2}  
% \end{equation}
Note that the $1/R$ dependence in the formulas above is fixed by the scale invariance of the CFT, and the proportionality with $C$ in equations \eqref{entropy}-\eqref{J} is due to the large-$C$ limit of the   CFT.  In what follows we shall make use of these variables to analyze the different  phases in the various  thermodynamic ensembles in the dual CFT.

\section{Thermodynamic  ensembles in the dual CFT}
\label{sec3}

 In this section we study the phase behaviour of different ``(grand) canonical" thermodynamic ensembles in the CFT, for thermal states that are dual to Kerr-AdS black holes. There are in principle eight   grand canonical ensembles in the CFT, since at fixed temperature there are three pairs of conjugate thermodynamic variables, namely $(\tilde \Omega, J)$, $(p, \mathcal V)$ and $(\mu, C)$.
In this paper we concentrate on the following three ensembles that feature interesting phase 
 behavior.
 %admit We find that only the following three   ensembles  have interesting phase behaviour; 
 We denote the associated free energies of the ensembles respectively as $F$, $W$ and~$G$:
    \begin{equation}
     \begin{aligned}  
    &\text{``canonical"} \quad  (J, {\cal V}, C): \qquad &&F \equiv E - TS   =\tilde \Omega  J +\mu C \,,\\
      &\text{``grand canonical"} \quad  (\tilde \Omega, {\cal V},C)\,: \qquad &&W \equiv E- TS - \tilde \Omega J =  
      \mu C\,,\quad \\
       &\text{``novel"} \quad  (J, {\cal V}, \mu)\,: \qquad &&G \equiv E - TS-     \mu C=  \tilde \Omega J\,, 
     \end{aligned}
 \end{equation}
 %\tcr{ {\bf why don't we study $\tilde\Omega, {\cal V}, \mu$ ensemble? Is it not interesting?}}
where, to obtain the second equalities, we  
have used the Euler equation \eqref{j2}.
%\begin{equation}
%F=E - TS \qquad 	W = E - TS- \tilde \Omega J \qquad \qquad G= E - T S - \mu C
%\end{equation}
%For the last two ensembles we used the Euler %equation \eqref{j2} to simplify the expression for the free energies.
In each case we shall also study the associated heat capacity, which gives a measure of thermodynamic stability of the system. We shall denote these as
\begin{equation}
    \mathcal{C}_{\chi} \equiv T \left (\frac{\partial S}{\partial T}\right)_{\chi}  \,,\qquad \chi \in \{(J, {\cal V}, C), (\tilde \Omega, {\cal V},C),  (J, {\cal V}, \mu)\}\,.
\end{equation}
Explicit expressions for the heat capacity in the latter two ensembles can be found below, while we omit the expression for $\mathcal{C}_{\chi}$ in  the first ensemble because it is too lengthy. The characteristic features of  $\mathcal{C}_{\chi}$ for the three ensembles are displayed below. 
%Figs.~\ref{}
%main features are  displayed in the various $\mathcal{C}_{\chi}$ plots.

 Before moving on, we would like to point out one recent related study in the literature. In \cite{Gao:2021xtt}, the authors studied the bulk thermodynamics of the  rotating AdS black holes in $d=4$ in the slowly rotating limit while keeping $\delta L=0$ (referred to as ``restricted phase space''). This is essentially equivalent to our fixed $\mathcal{V}$ ensembles (with $\omega =1$).  %\tcp{\bf This cannot be right because our $\cal V$ is independent of $L$. Does it correspond to fixed $C$ ensemble?}. \rbm{From (1.5), $\cal V$ depends on $L$.}
 However, we are here interpreting the results from the point of view of the boundary CFT instead of the bulk gravity theory, and without going to the slowly rotating limit. The interested reader is thus invited to visit \cite{Gao:2021xtt} to see how some of our results can be interpreted from the bulk perspective.

In what follows, the values for the dimensionful quantities $\{F,W,G,T,\tilde\Omega,\mu\}$ (including in all figures) will always be understood to be given in units $1/\ell$, where $\ell$ is an arbitrary constant length scale. Similarly, the values of $\mathcal{V}$ will be give in units of $\ell^{d-2}$. Furthermore, we note from the expressions \eqref{entropy}-\eqref{mu} that the scale $R$, and hence $\mathcal{V}$, does not affect the qualitative thermodynamic behaviour of the system. Hence we set $\mathcal{V} = 1$ in all illustrative figures.

\subsection{Canonical ensemble: $F(T,J, \mathcal V,C)$}
The canonical 
(fixed $T,J,P$) ensemble
%fixed $(J, \mathcal V,C)$ ensemble 
has been well studied in the framework of black hole chemistry in the bulk   \cite{Altamirano:2013ane,Gunasekaran12}. 
%where $\Lambda$ was allowed to vary but $G$ was kept fixed. 
In those studies it was found that the smallest dimension that  displays interesting phase behaviour is $d=4$, where a \textit{Van der Waals like phase transition} takes place between small and large rotating black holes. While the behaviour in $d=5$ is qualitatively similar to $d=4$, black holes in dimensions $d\geq 6$ can undergo \textit{reentrant phase transitions}, which are absent in the lower dimensions. In this section  we study the analogous ensemble in the CFT in $ d=4$ and $d=6$, which is given by not only fixing the angular momentum, but also holding fixed the volume and central charge.

%\subsubsection{Free Energy $F$}

The relevant free energy in the fixed ($J,\mathcal  V, C$) canonical ensemble is
\begin{equation}
\label{F}
    F %\equiv E - TS 
    = \frac{C x^{d-5}}{R \left(z^2-1\right)^2} \left(x^4 \left(3 z^2-1\right)+x^2 \left(z^2+1\right)^2-z^4+3 z^2\right)\,.
\end{equation}
We note that $z$ in the above should be viewed as a function of $C$ and $J$ (and $x$),  which can be obtained by inverting the expression~\eqref{J} for $J$. While the actual solution is too long to be included here, we note %it is clear 
from~\eqref{J} that $z$ is a function of $x$ and of the ratio 
\be 
\kappa\equiv J/C\,.
\ee
Together with the expressions~\eqref{T} and~\eqref{F} for $T$ and $F$ respectively, this implies that $T$ and $F/C$ are functions of $(x, \mathcal{V}, \kappa)$. We shall see the implications of this below.

\subsubsection{$d=4$: Swallowtail criticality}
\begin{figure}[b]
    \centering
    \includegraphics[scale=0.8]{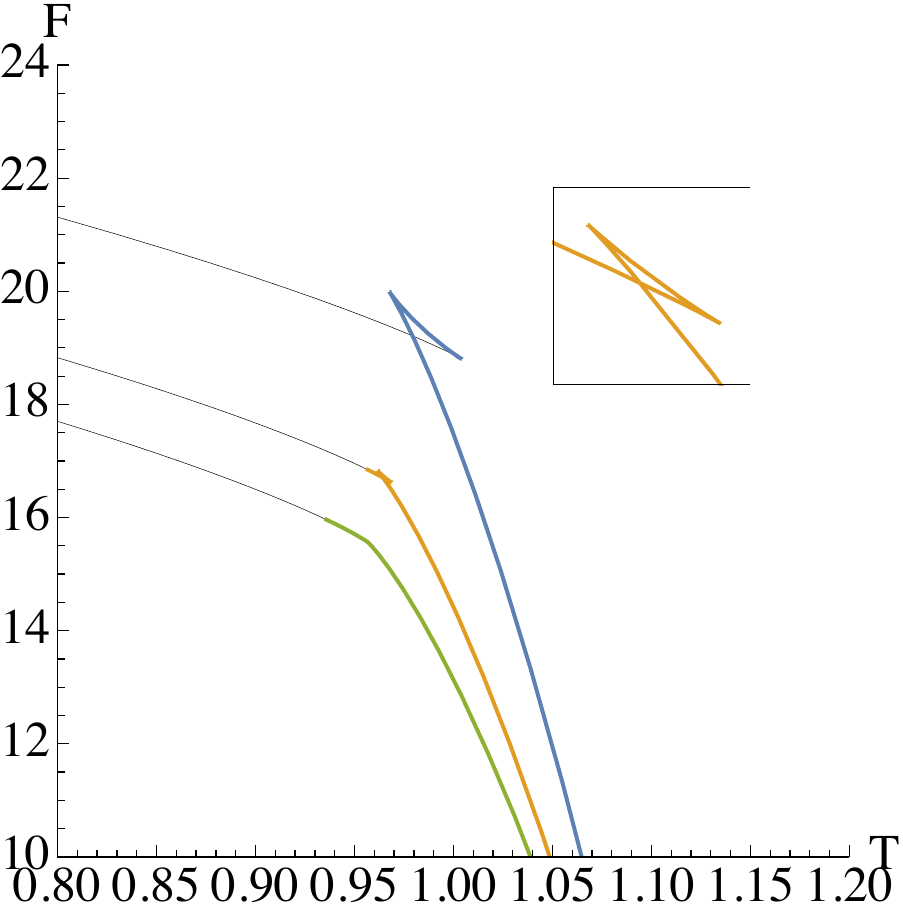}
    \includegraphics[scale=0.8]{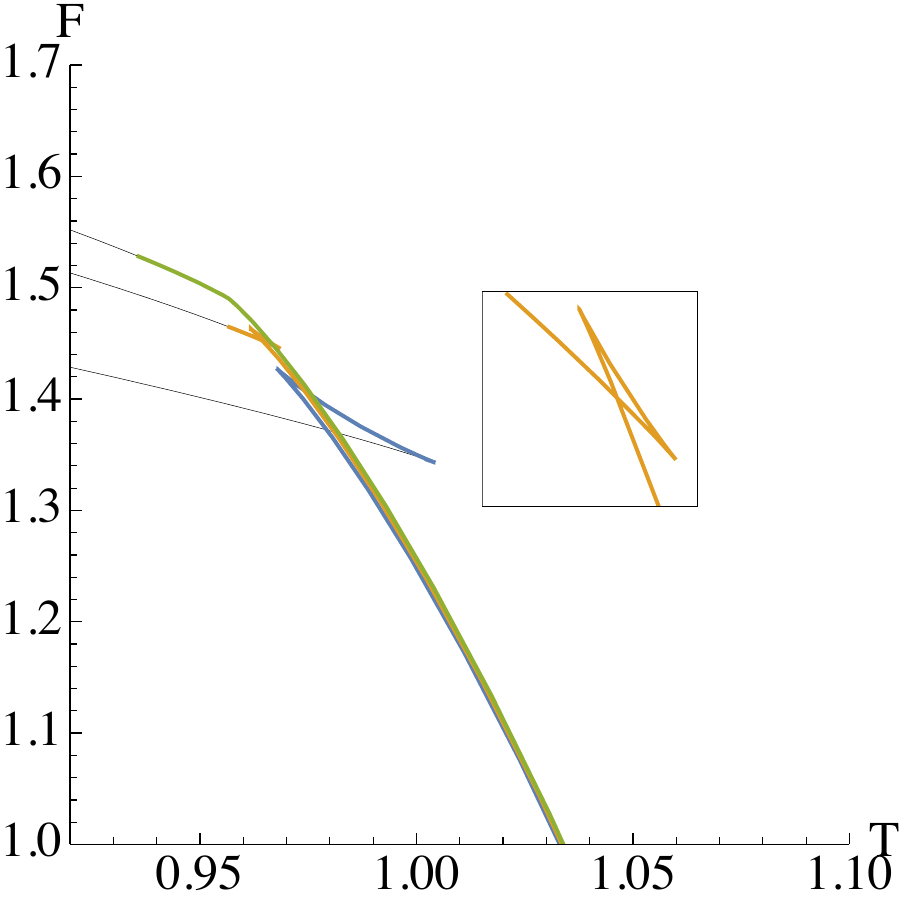}
    \caption{$F-T$ diagram of fixed $(J, \mathcal V,C)$ ensemble for $d=4$, $\mathcal V=1$. The free energy against temperature diagram is plotted here at $J=1$ and various fixed $C$ on the left, and $C=1$ at various fixed $J$ on the right.  \textbf{Left}: $J=1$, $C=14$ (blue), $C=11.5$ (yellow), $C\approx 10.45$ (green, critical). \textbf{Right}: $C=1$, $J=1/14$ (blue), $J=2/23$ (yellow),   $J \approx 0.0957$ (green, critical).  The plot is swallowtail-shaped below a critical $J/C$ ratio (see main text).  Note that each curve in the left diagram is simply ``stretched'' along the $F$ axis as compared to its counterpart on the right. The black portions of the curves denote the solutions with 
    $\Omega L > 1$, where
    superradiant instabilities are present in the bulk. The insets show zoom-ins of the yellow swallowtail.
 % \rbm{Will a different value of $J$ push the superradiant part leftward?  It would be nice if the swallowtails didn't intersect on a black curves. Wan: Done.}  
    }
    \label{fig:FTd4}
\end{figure}

Let us begin by considering the $F-T$ diagram in $d=4$ dimensions (see Fig. \ref{fig:FTd4}).  On the left, %  plot, 
the value of $J$ is kept fixed at $J=1$, and the different curves correspond to varying $1/C$; the roles of $J$ and $1/C$ are swapped on the right. These diagrams are reminiscent of the analogous diagrams for Van der Waals (VdW) fluids -- the blue and yellow curves of these figures resemble the shape of a swallowtail. However these exist only \textit{below} some \textit{critical value},
$\kappa<\kappa_{4,crit}$, where we derive the numerical value for $\kappa_{4,crit}$ below in equation \eqref{k4crit}. 
%of the ratio $\kappa \equiv J/C$. We derive the numerical value of $\kappa_{4,crit}$ below.

Each swallowtail consists of three piecewise smooth branches. Since the entropy \eqref{entropy} is an increasing function of $x$, we shall call the gently sloping branch starting from $T=0$  the \textit{low entropy (LE)} branch, as the value of $x$ 
is the smallest on this branch. The steep, negatively sloped branch extending to $F\rightarrow -\infty$ is called the \textit{high entropy (HE)} branch  as $x$ is largest on this branch. The intermediate branch joining these two has intermediate $x$ values, and is called the \textit{intermediate entropy (IE)} branch. For any swallowtail curve, starting at high $T$, the branch that minimizes the free energy and is thus thermodynamically favoured, is  initially given by the HE branch. However at the self-intersection temperature, the $F-$minimizing branch changes to the LE branch. Standard thermodynamic arguments imply that the system, which in our case is a thermal CFT, undergoes a first-order phase transition at this temperature between the HE and LE phases. We note that these two phases have positive heat capacity $\mathcal{C}_{J,\vol,C}$ and hence are thermodynamically stable, while the IE branch has $\mathcal{C}_{J,\vol,C}<0$, 
as can be seen in Fig.~\ref{fig:HCFT}.
% $F-$minimizing, and thus, the thermodynamically favoured, branch i
%see Fig. \ref{fig:HCFT}. 
This phase transition becomes second order at the \textit{critical point} where the values of $(J,C,T)$ are such that $J/C=\kappa_{4,crit}$ and  $T=T_{crit}$. The $F-T$ curve (green) displays a kink at this critical point. Above the critical value for $J/C$ the free energy curves will be smooth and single valued. This behaviour is  typical swallowtail criticality,  which is also present for charged AdS black holes \cite{Chamblin:1999tk}. 

\begin{figure}[b]
    \centering
    \includegraphics[scale=0.8]{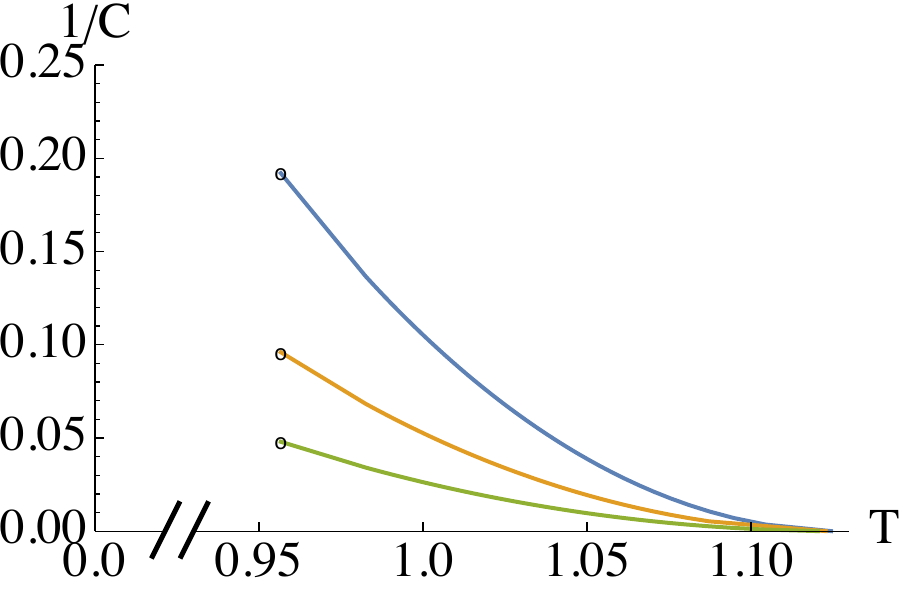}
    \includegraphics[scale=0.8]{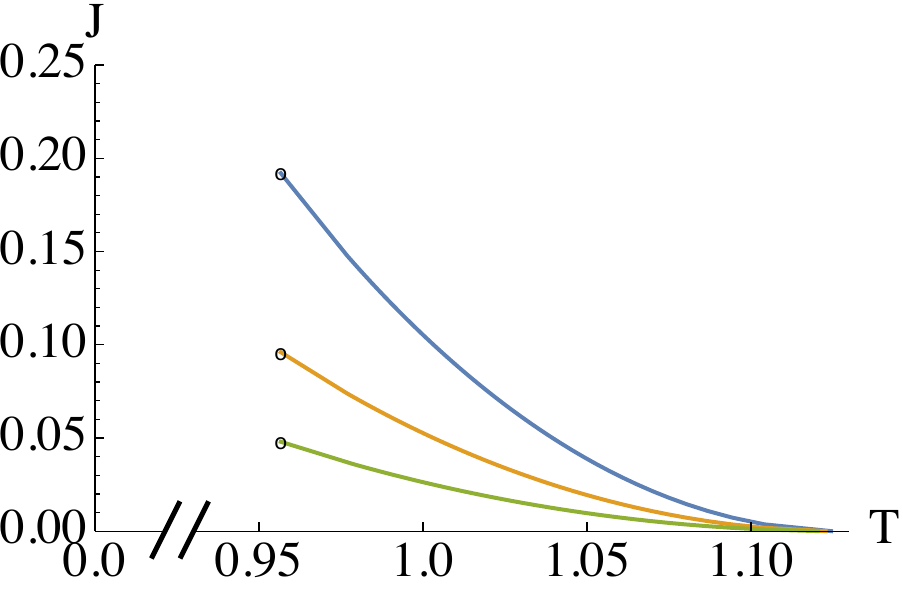}
    \caption{Co-existence diagram in $d=4$. Each of the curves displayed here is a line of first-order phase transitions for different parameter values. \textbf{Left}: $J=1/2$ (blue), $J=1$ (yellow), $J=2$ (green) for $\mathcal{V}=1$. \textbf{Right}: $1/C=1/2$ (blue), $1/C=1$ (yellow), $1/C=2$ (green) for $\mathcal{V}=1$. For each of these parameter values, the line of first-order phase transition separates the  low-entropy (LE) phase, lying to the left of the curve, and the  high-entropy (HE)  phase, lying to the right. Each line ends at a critical point, denoted by an open circle, where the phase transition becomes second order.}
    \label{fig:FTphase4d}
\end{figure}

 A notable distinction from the standard first-order transitions seen for AdS black holes is that the smaller black hole branch has  superradiant instabilities. In other words, as temperature decreases the first-order transition is from a large black hole to a small one with $\Omega L > 1$.  This branch is presumably replaced with a branch of stable small black holes with some kind of scalar hair \cite{Cardoso:2006wa, Kim:2023sig}. We shall not pursue this issue further.

The co-existence phase diagrams for these transitions are plotted in Fig. \ref{fig:FTphase4d}. Each curve on these diagrams is a line of first-order phase transitions that terminates at a critical point denoted by open circles. The HE phase lies to the right of the curves while the LE phase lies to the left of the curves. The two phases become indistinguishable above the critical points. Notice that the left and right diagrams in Fig. \ref{fig:FTphase4d} are identical. This is due to the above mentioned dependence of $T$ and $F/C$ on only the ratio $J/C$ instead of on $J$ and $C$ independently. As a result, varying $1/C$ at fixed  $J$ has the same ``thermodynamic effect'' as varying $J$ at fixed $1/C $. In particular, looking at Fig. \ref{fig:FTd4}, we see that each curve in the left diagram is only stretched along the $F$ axis as compared to the corresponding curve on the right, but the phase transition temperatures are identical. This explains the identical diagrams in Fig. \ref{fig:FTphase4d}.

We also note that the $T$-intercepts of the co-existence lines all occur at the same value of $T$. This temperature is given by the Hawking--Page transition temperature $T_{\mbox{\tiny HP}}$ at $J=0$ which can be obtained by solving for $x$ in~\eqref{F} i.e., $F(x, z=0)=0$,  for which one gets the solution $x=1$ (or $r_h = L$). Substituting this and $z=0$ into the expression for the temperature $T$ then gives the value 
\be 
T_{\mbox{\tiny HP}} = \frac{1}{\pi R}\,.
\ee
The thermodynamic behaviour in $d=5$ is qualitatively similar to that in $d=4$, displaying in particular the same swallowtail criticality. We shall not consider this case further.

\subsubsection{$d=6$: Reentrant phase transition}

\begin{figure}
    \centering
    \includegraphics[scale=0.8]{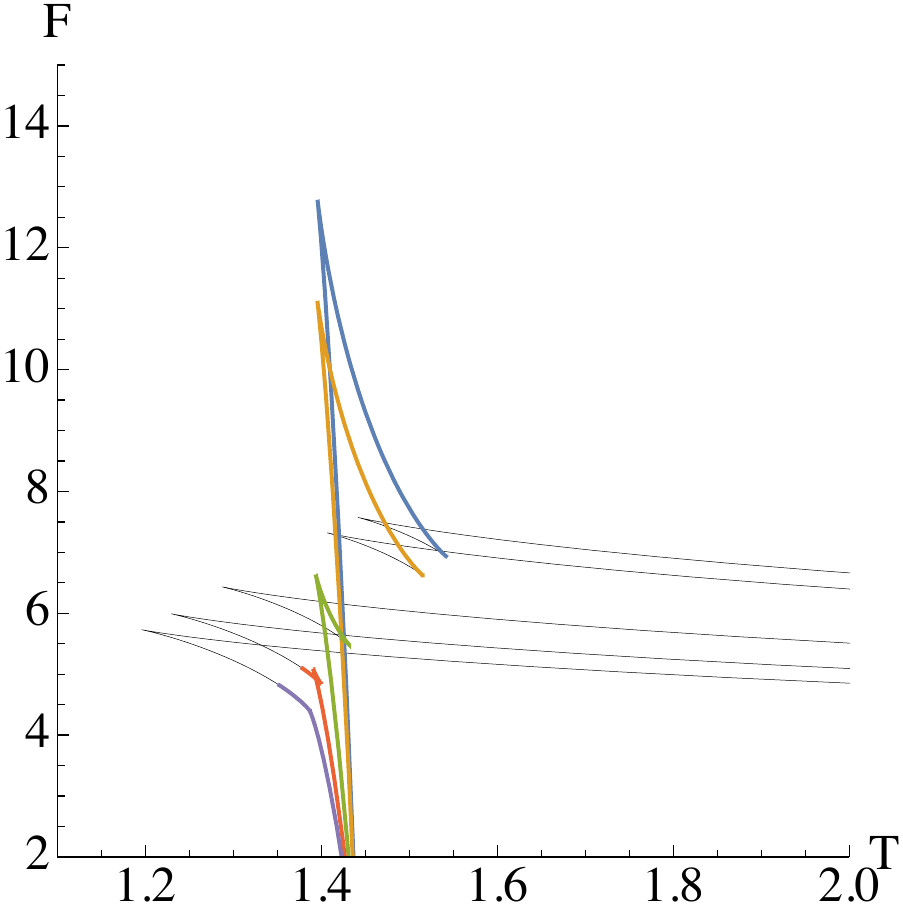}
    \includegraphics[scale=0.8]{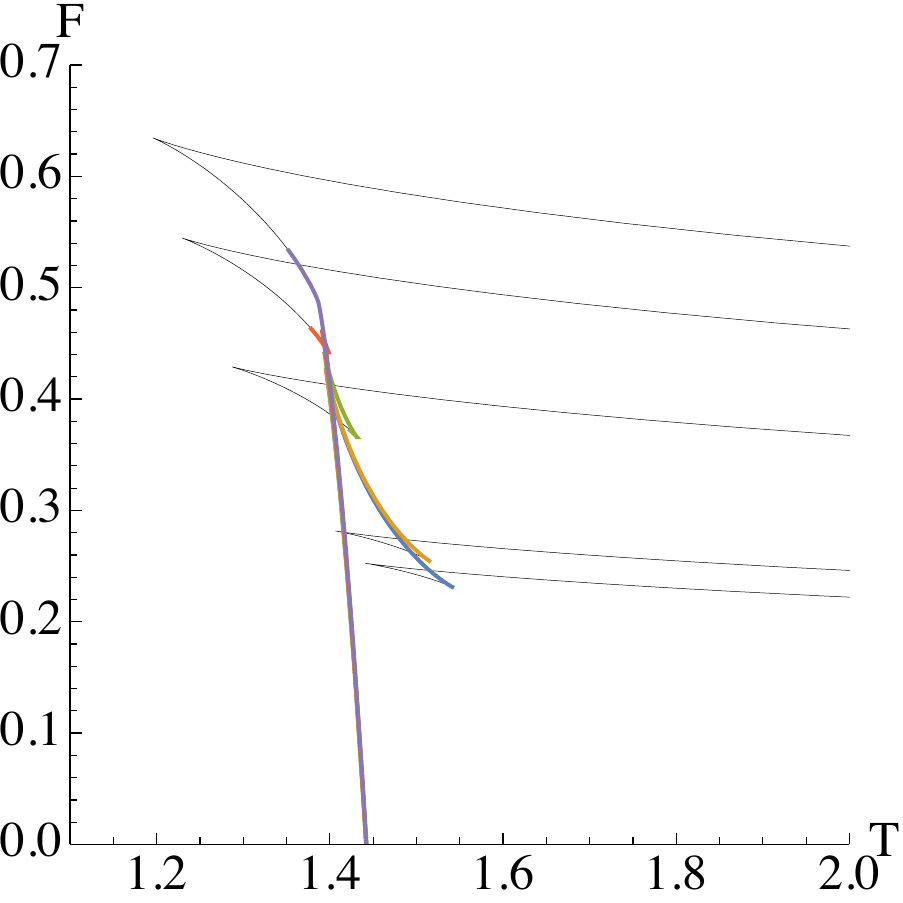}
    \caption{$F-T$ diagram of fixed $(J, \mathcal V,C)$ ensemble for $d=6$, $\vol=1$. \textbf{Left}: $J=1$, $C=30$ (blue), $C=26$ (yellow), $C=15$ (green), $C=11$ (red), $C\approx 9.03$ (purple, critical). \textbf{Right}: same values as left, with $J\leftrightarrow 1/C$. As in the $d=4$ case, the left diagram is simply a stretch of the right diagram along the $F$ axis.  The black portions of the curves denote the solutions with 
    $\Omega L > 1$, where
    superradiant instabilities are present in the bulk.  
Note that for the yellow and green curves there are HE to LE first-order transitions as the temperature decreases; however the LE branches are (partly) superradiant.
    }
    \label{fig:FTvC}
\end{figure}

\begin{figure}
    \centering
    \includegraphics[scale=0.8]{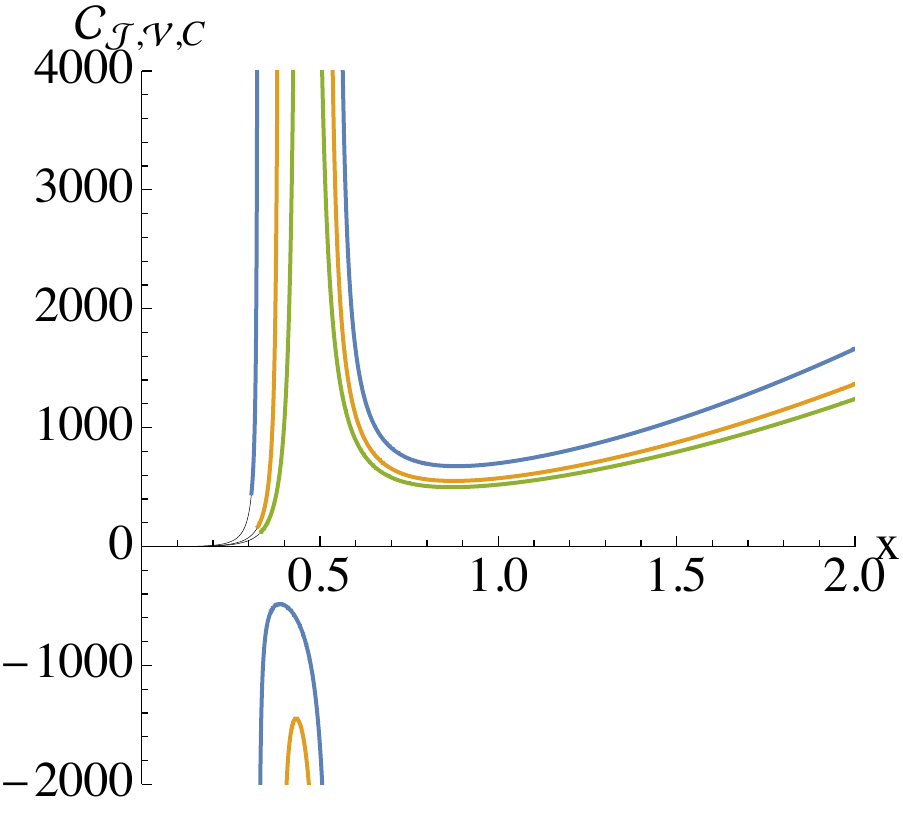}
    \includegraphics[scale=0.8]{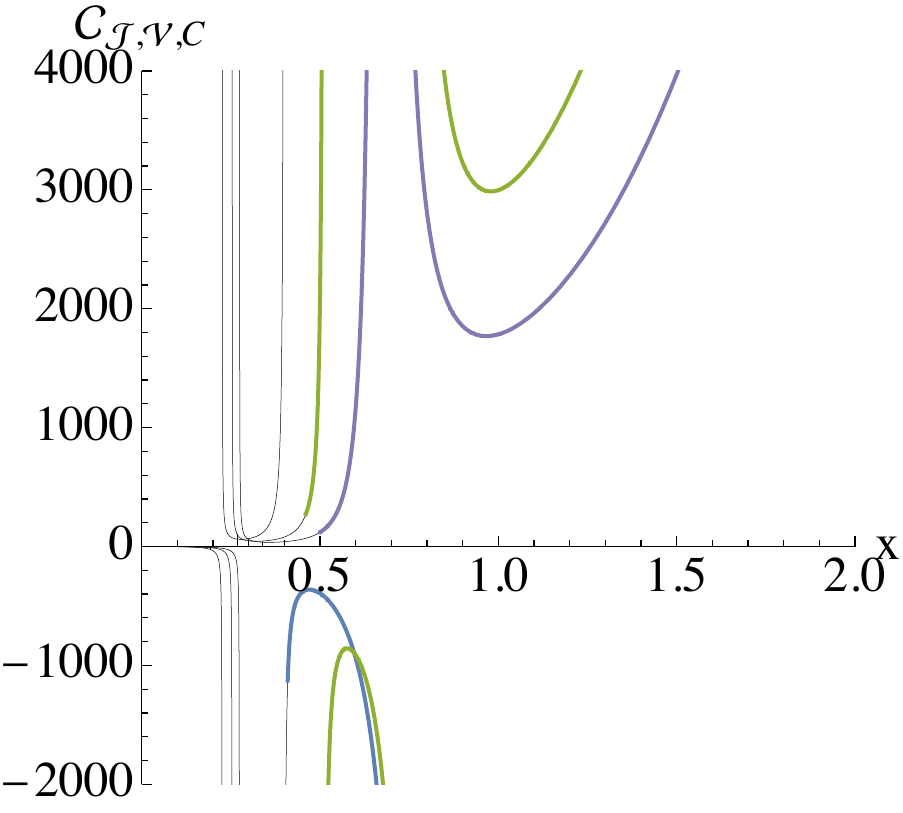}
    \caption{Heat Capacity $\mathcal{C}_{J,\mathcal{V},C}$ against $x$. The parameters used here are the same as those  in Figs. \ref{fig:FTd4} and \ref{fig:FTvC}, respectively. \textbf{Left:} $d=4$, $J=1$ and $C=14$ (blue), $C=11.5$ (yellow), $C\approx 10.45$ (green, critical). The first two curves obey  $\kappa<\kappa_{4,crit}$ and consist of three piecewise continuous segments, corresponding to the three branches of the $F-T$ diagram: the LE, IE and HE phases. The LE phase corresponds to the segment with smallest $x$, having $\mathcal{C}_{J,\mathcal{V},C}>0$; the IE phase has intermediate $x$ and $\mathcal{C}_{J,\mathcal{V},C}<0$; the HE phase has largest $x$ and $\mathcal{C}_{J,\mathcal{V},C}>0$. \textbf{Right:} $d=6$, $J=1$, $C=30$ (blue), $C=15$ (green), $C\approx 9.03$ (purple, critical). Here each curve with $\kappa<\kappa_{6,crit}$ consists of four piecewise continuous segments, in correspondence with the four branches in the $F-T$ diagram. Phase transitions take place between the two segments with $\mathcal{C}_{J,\mathcal{V},C}>0$, corresponding to what was referred to as the LE (segment with relatively smaller $x$) and HE phases in the main text. Black lines correspond to superradiant phases in the bulk.}
    \label{fig:HCFT}
\end{figure}

 The $F-T$ diagram for $d=6$ is displayed in Fig. \ref{fig:FTvC}.  For $\kappa< \kappa_{6, crit}$, 
 %Now, below the critical $J/C$ value $\kappa_{6, crit}$, 
 each curve consists of four branches, allowing for more elaborate phase behaviour. 
 In each case, one branch corresponds to a high-entropy state with positive heat capacity, and there is a low-entropy branch with negative heat capacity. The other two intermediate entropy branches have either negative or positive heat capacity, as can be seen on the right diagram of Fig.~\ref{fig:HCFT}. Compared to the $d=4$ case, the  low-entropy branch with  negative heat capacity is novel. 
 
 %; there is also another   intermediate entropy state having positive heat capacity. The other two segments have negative heat capacity.  
 
\begin{figure}
    \centering    \includegraphics[scale=1]{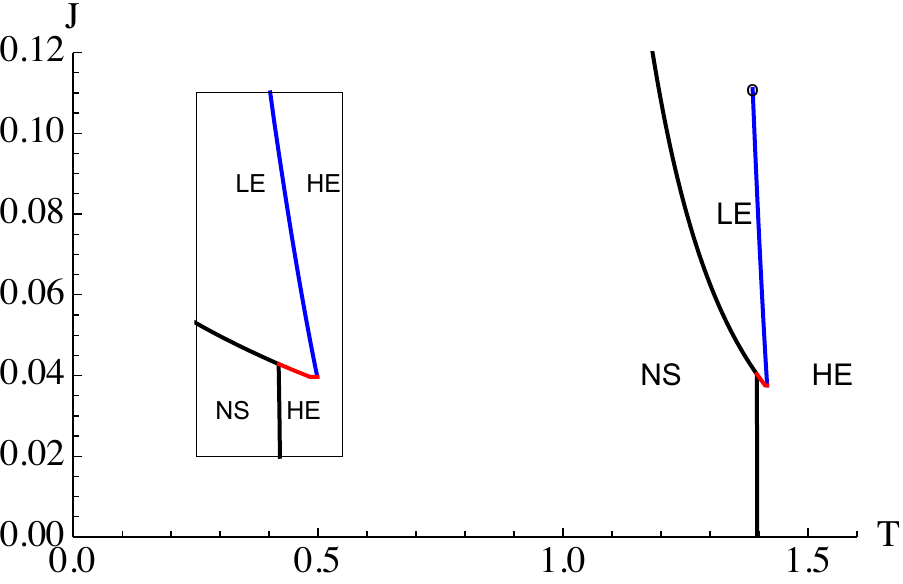}
    \caption{Phase diagram in fixed $(J,\mathcal{V},C)$ ensemble, for $d=6$ and  $C=\mathcal{V}=1$, showing a (blue) line  of first-order phase transitions between high-entropy (HE) and low-entropy (LE) phases.
    The inset shows a close up around the  (red) line of zeroth-order phase transitions between the  LE and HE  phases. The system can in fact undergo a HE-LE-HE reentrant phase transition as we lower $J$ at a fixed temperature admitting zeroth-order phase transition. The region lying to the left of the black lines 
    is a region where no solutions (NS) are  possible. 
     }
    \label{fig:FTphase6D}
\end{figure}

 In the right diagram of  Fig.~\ref{fig:FTvC}, for sufficiently small $J$ 
 (the blue $J=1/30$ curve) there is a cusp in the free energy diagram accompanied by
 an inverted swallowtail at higher temperatures, and so this case has only one phase. However for larger $J$ the situation changes:  the  $J=1/26$ (yellow) case  implies a \textit{reentrant phase transition}, where the inverted swallowtail now intersects the steep HE curve. Here the 
 $F$-minimizing branch changes from the HE branch at (slightly) higher $T$ to the LE branch before jumping back to the HE branch at some lower $T$. This last phase shall still be called the HE phase, though it has
 lower entropy than
 the original HE segment.
 This second transition is accompanied by a jump in the free energy value and is hence a \emph{zeroth-order phase transition} (ZOT). The line of ZOTs is depicted by the red line in the $J-T$ phase diagram of Fig. \ref{fig:FTphase6D}. As in the $d=4$ case, the $1/C-T$ phase diagram is qualitatively similar to Fig. \ref{fig:FTphase6D} and is omitted. 

 For larger $J$ (or smaller $C$) the situation changes further. The  $J=1/15$ (green) case is an almost-star-shaped curve. As in the $d=4$ case, the steep negatively sloping branch has the largest $x$ values. At high $T$, this HE branch initially minimises $F$ but as $T$ decreases there will be a first-order phase transition when this branch intersects the LE branch. These two branches have positive heat capacities, whereas the other two branches have negative heat capacities, as shown in Fig. \ref{fig:HCFT}. 
 
 For larger $J$ we then have the familiar swallowtail corresponding to a first-order transition between the HE and LE states, up to the critical value  (purple),
 with   $J=0.11$. 
 This behaviour is also visible in the phase diagram in Fig. \ref{fig:FTphase6D}, where we continue to see a line of first-order phase transitions, ending at a critical point depicted by an open circle. However, from the $F-T$ diagram, we see that the curves now do not intersect the $T=0$ axis unlike the $d=4$ case. This gives rise to an additional region in the phase diagram, lying to the left of the black line, where no solution exists. The black line intersects the $T-$axis at $T=\frac{\sqrt{15}}{2\pi R}$. This is the temperature at $J=0$ below which there is no solution (NS). From the $F-T$ diagram, we can observe that this happens at the minimum of $T(x)$, i.e.  $\partial T/\partial x|_{z=0} = 0$, which occurs at $x=\sqrt{d-3}/\sqrt{d-1}$. The same qualitative phase behaviour was found in $d=7$ and we did not find any new behaviours for higher dimensions.

\subsubsection{Critical points}

We now derive the numerical values of the critical point. In any $d$, the critical point is characterised by:
\begin{equation}
\label{critpt}
\frac{\partial C}{\partial \mu}\Big |_{T,J,\mathcal V} = 0=\frac{\partial^2 C}{\partial \mu^2}\Big |_{T,J,\mathcal V} \,.
\end{equation}
To this end, we first solve~\eqref{T} for $z(T,R,x)$, and substitute this solution into~\eqref{J} and~\eqref{mu} to obtain $C(J,x,z( T,R,x))$ and $\mu(R,x,z(T,R,x))$. This gives:
\begin{align}
    C &= \frac{J x^{1-d} \left(-d x^2-d+4 \pi  R T x+x^2+5\right)^2 \sqrt{-\frac{x^2 \left(d x^2+d-4 \pi  R T x-x^2-3\right)}{d x^2+d-4 \pi 
   R T x-3 x^2-5}}}{4 \left(d x^2+d-4 \pi  R T x-x^2-3\right)}\,, 
   \\
   \mu &= \frac{2 \left(x^2-1\right) x^{d-3}}{(d-1) x^2+d-4 \pi  R T x-5} \,.
\end{align}
From this we then solve numerically \eqref{critpt} for the critical point as the root of some polynomial equation whose degree depends on $d$. In $d=4$ the critical point, which is $C$- and $J$-independent, is
\begin{equation}
    T_{crit} R \approx 0.26987 \,,\qquad x_{crit}  \approx 0.45882 \,,
\end{equation}
corresponding to 
\begin{equation} \label{k4crit}
    \kappa_{4,crit} \approx 0.095732\,.
\end{equation}
It is interesting to compare this to the result in \cite{Gao:2021xtt}, which found that $\kappa_{4,crit}\approx 0.096424$ in the slowly rotating limit.

In $d=6$ the critical point is,
\begin{equation}
    T_c R \approx 0.612275 \,,\qquad x_c  \approx 0.679425 \,,
\end{equation}
corresponding to 
\begin{equation} \label{k6crit}
    \kappa_{6,crit} \approx 0.1107\,.
\end{equation}
% In $d=5$ the critical point is,
% \begin{equation}
%     T_c R \approx 0.4455 \,,\qquad x_c  \approx 0.597 \,,
% \end{equation}
% corresponding to 
% \begin{equation}
%     \frac{J_{crit}}{C_{crit}} \approx 0.108\,.
% \end{equation}

% Heat capacity
% \begin{equation}
%     \mathcal{C}_{\mathcal{J},\mathcal{V},C} = T\frac{\partial S}{\partial T}|_{\mathcal{J},\mathcal{V},C}
% \end{equation}
% analytic expression too long and not instructive. Same qualitative result as in the bulk, see Fig. \ref{fig:HCFT}.

\subsection{Grand canonical ensemble: $W(T, \tilde \Omega, {\cal V}, C)$}

Next we study the fixed ($\tilde \Omega, \mathcal V, C$) ensemble, usually referred to as the ``grand canonical ensemble''. The free energy in this ensemble can be expressed as
\begin{equation} \label{Wfreeenergy}
    W %&=&\tcr{E-TS-\tilde \Omega J}\nonumber\\ 
   =\frac{Cx^{d-5}\left(x^2-1\right)  \left(x^2+z^2\right)}{R
   \left(z^2-1\right)} = \frac{C\left(\tilde{\Omega} R \left(z^2+1\right)-2 z\right) \left(\frac{\sqrt{z-\tilde{\Omega} R z^2}}{\sqrt{\tilde{\Omega} R-z}}\right)^{d-1}}{R z
   (\tilde{\Omega} R z-1)^2}\,,
\end{equation}
where we have used
\begin{equation}
    x = \frac{\sqrt{z-\tilde{\Omega} R z^2}}{\sqrt{\tilde{\Omega} R-z}}
\end{equation}
to obtain the second equality. Note that the condition $x>0$ restricts the physical parameter values to $0<z<\tilde \Omega R$, if $0< \tilde\Omega R \leq 1$, and to $0<z<1/(\tilde \Omega R)$, if $ \tilde\Omega R > 1$. In particular, as we shall see shortly, the point $(\tilde\Omega R,z) = (1,1)$ corresponds to a  `transition point'. 

Further, we can obtain the expression for the temperature in terms of the variables $(\tilde{\Omega},R,z)$ in the same way, which yields
\begin{equation}
    T=\frac{\tilde{\Omega} R \left(d-3-2 \tilde{\Omega} R z-(d-3) z^2\right)+2 z}{4 \pi  R \sqrt{z(\tilde{\Omega} R-z)(1-\tilde{\Omega} R z)}}\,.
\end{equation}
 This allows one to plot the $W-T$ diagram parametrically. We refer the reader to App.~\ref{appA} for the discussion of bulk thermodynamics in the grand canonical ensemble, for comparison.

\begin{figure}
    \centering
    \includegraphics[scale=0.8]{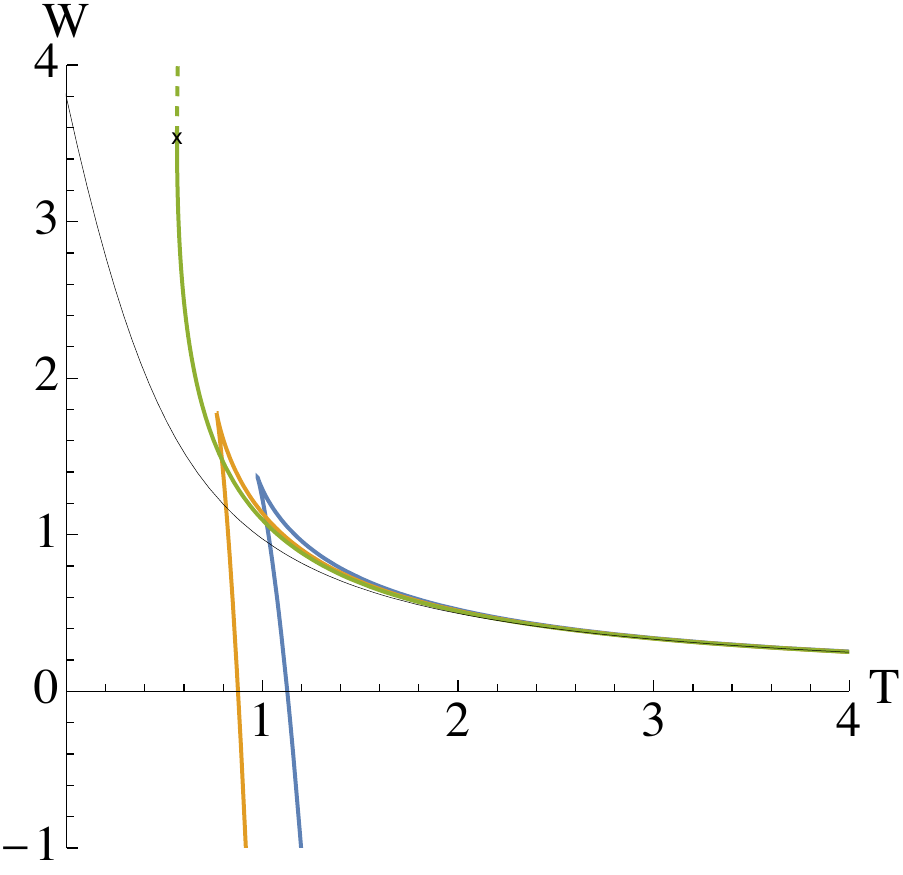}
    \hspace{0.5cm}
    \includegraphics[scale=0.8]{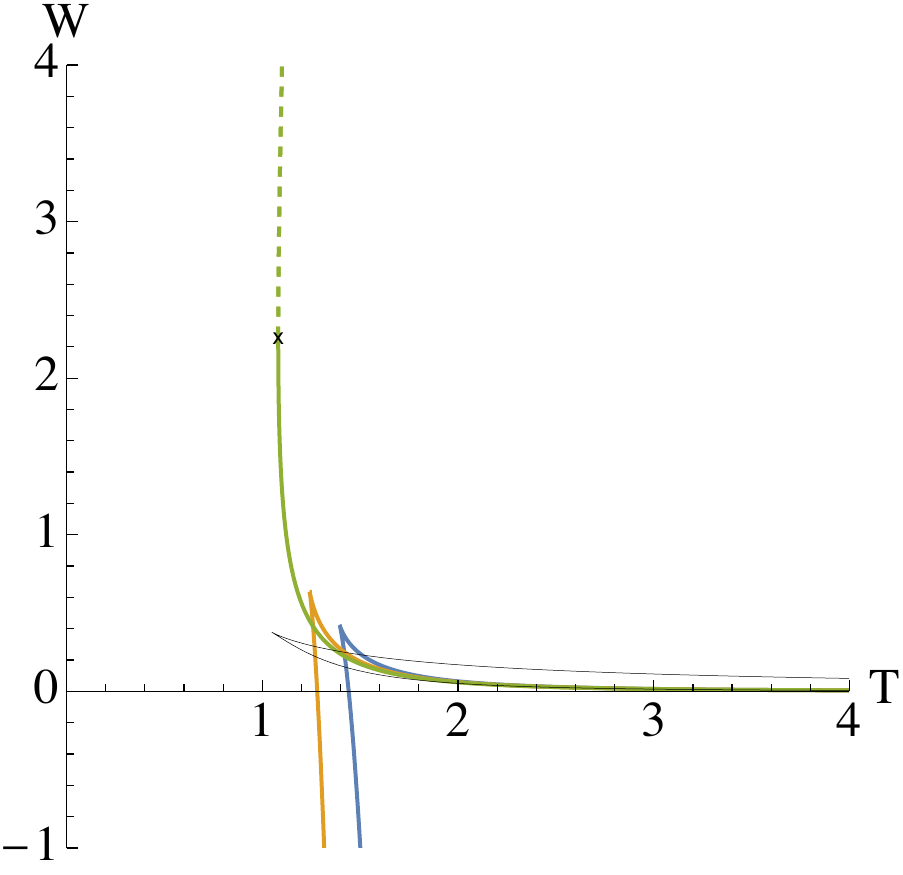}
    \caption{$W-T$ diagram of fixed $(\tilde\Omega,\vol,C)$ ensemble for $\vol=1=C$, $d=4$ (left), and $d=6$~(right). The curves correspond to $\tilde\Omega R= 1/20 $ (blue), $\tilde\Omega R=5/6 $ (yellow), $\tilde\Omega R= 1$ (green) and $\tilde\Omega R= 3/2 $ (black). For $\tilde{\Omega} R=1$, the physical part of the figure corresponds to the solid line, while the dashed part has $z>1$.
    The black lines correspond to superradiant black holes in the bulk.} 
    \label{fig:WT}
\end{figure}

\subsubsection{(De)confinement phase transition}

\begin{figure}
    \centering
    \includegraphics[scale=0.7]{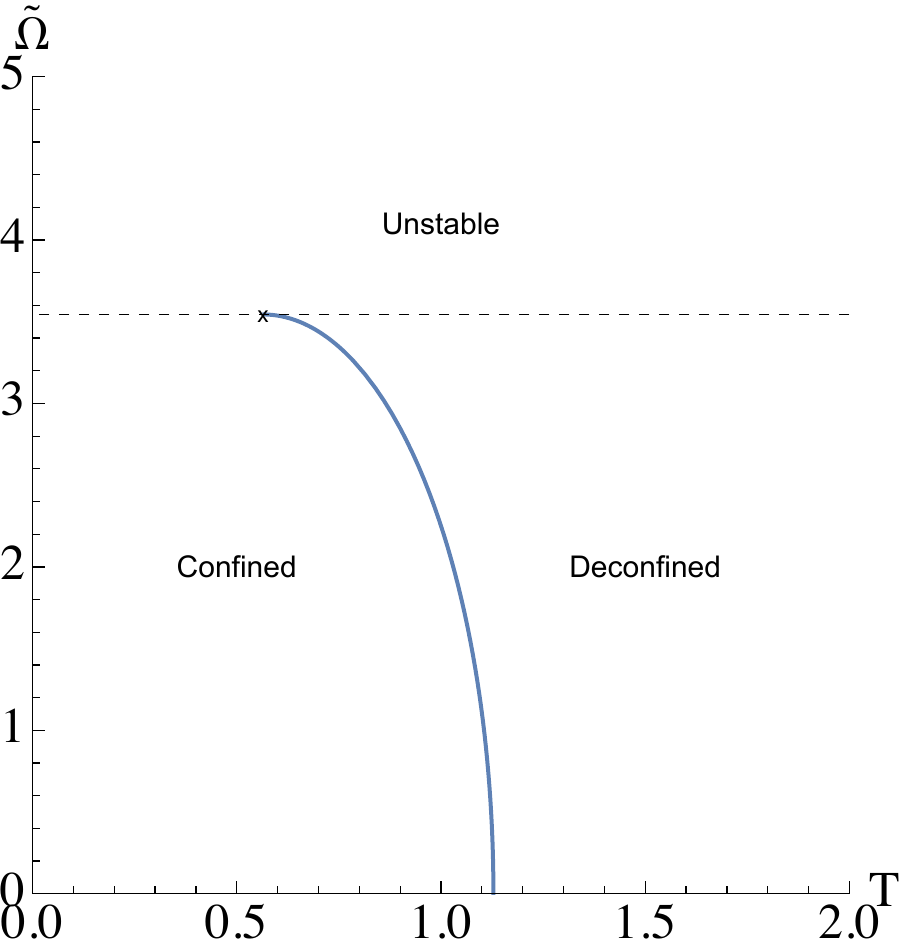}
        \hspace{0.5cm}
    \includegraphics[scale=0.7]{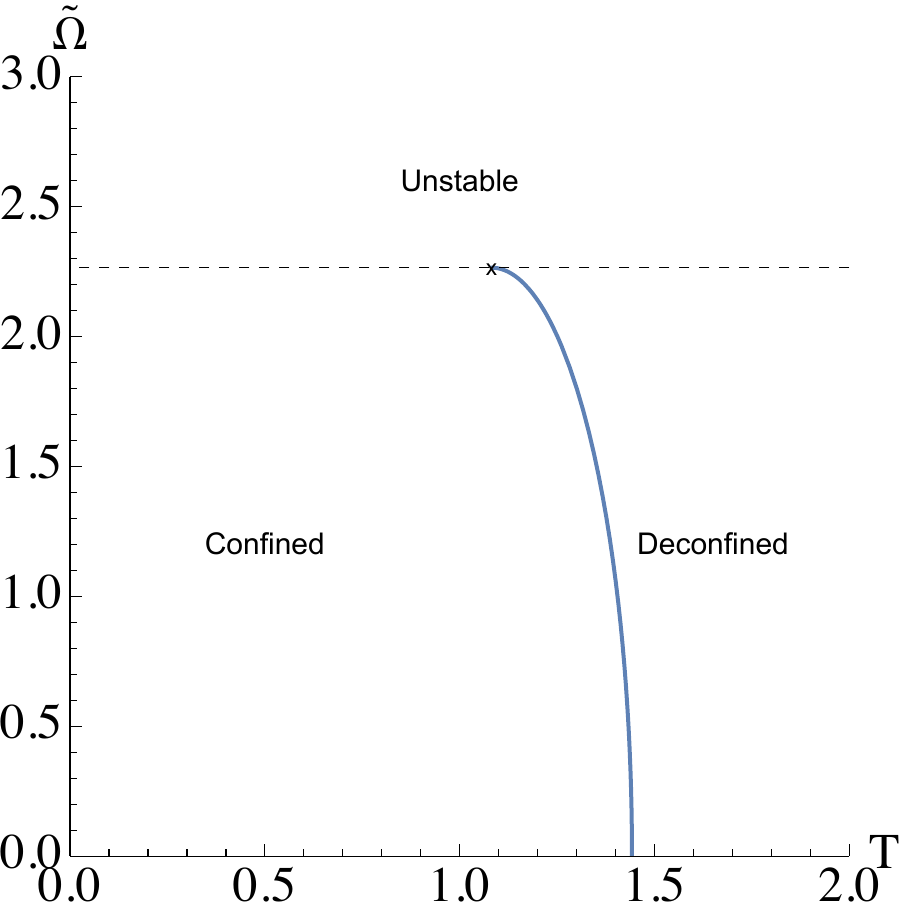}
    \caption{Co-existence diagram for $\tilde \Omega$ vs. $T$. The parameters used here are $C=1=\vol$, $d=4$ (left), and $d=6$ (right). For $\tilde\Omega R<1 $,  a first-order phase transition occurs across the co-existence line separating the confined and deconfined phase. 
    Contrary to the canonical ensemble, the coexistence line no longer terminates at a critical point. Rather, an `unstable region' (subject to superradiant instabilities in the bulk) develops for $\tilde \Omega R\geq 1$. 
    %\MR{'cross X' in both figures at the intersection of the dashed and solid lines, i.e. at the endpoint of the solid line added.}
      %However for $\tilde\Omega\geq 1/R$, only the confined phase is present.}
    }\label{fig:WTCoE}
\end{figure}

Let us first study the $W-T$ diagram, which is plotted in Fig. \ref{fig:WT} for $d=4$ (left) and $d=6$~(right). For values of $\tilde\Omega R<1$, the free energy vs. temperature diagram suggests that a first-order  (Hawking-Page like) phase transition  occurs   at the point where the curve cuts the $T-$axis at
\begin{equation}
    z  = \frac{1-\sqrt{1-R^2 \tilde\Omega ^2}}{R \tilde\Omega }\,,\qquad
    T = \frac{d-3+\sqrt{1-R^2 \tilde \Omega ^2}}{2 \pi  R}\,.
\end{equation}
This gives a line of first-order phase transitions on the $\tilde\Omega-T$ co-existence diagram, plotted in~Fig.~\ref{fig:WTCoE}    for $\tilde \Omega R \ge 0$. For $\tilde \Omega =0$ this agrees with  the standard Hawking-Page phase transition at temperature $T_{\text{HP}}=\frac{d-2}{2 \pi R}$, but for nonzero angular velocity we have an  entire line of phase transitions. We   labeled the phase at low temperature as the \textit{confined} phase, which is dual to thermal radiation in AdS, and the phase at higher temperature as the \textit{deconfined} phase, which is dual to a large black hole. The deconfined phase dominates the ensemble if  $W<0$ solutions on any given $W-T$ curve for $\tilde\Omega R <1 $, while the confined phase dominates if the curve satisifes $W>0$. The confined phase has $W=0$, since   we defined  $W$ in the bulk as   the free energy of the black hole system minus that of thermal AdS.

Furthermore, for $\tilde\Omega R > 1$  the free energy $W$ is  greater than zero and the curves no longer intersect the $T-$axis.
As a result, in this regime the $W-$minimizing phase is always  given by the confined phase (see also \cite{Hawking:1999dp}). When $\tilde \Omega R= 1$, the expressions for $W$ and $T$ reduce to
\begin{align}
   \tilde \Omega R =1 : \qquad   W=\frac{z^{(d-3)/2}}{R}\,,\qquad
    T= \frac{(d-3) (z+1)}{4 \pi  R\sqrt{z}}\,.
\end{align}
In particular, if $z=1$  we have $(T_\times,W_\times):=((d-3)/(2 \pi  R) ,1/R)$. This point is indicated by the cross (``x'') in Fig. \ref{fig:WT}. The range $z>1$ is plotted in Fig. \ref{fig:WT} using dashed lines to indicate that this is an unphysical region with $a>L$,  which corresponds to an overspinning black hole.   This range is not included in the co-existence diagram in Fig.~\ref{fig:WTCoE}. 

The behaviour here for the CFT dual to the Kerr-AdS family is markedly different from that dual to the analogous ensemble for the charged Reissner--Nordstr\"{o}m AdS family. The latter has been studied for example in \cite{Cong:2021jgb}, with fixed electric potential $\tilde \Phi$, in place of the fixed angular velocity $\tilde \Omega$ here. Like in the rotating case, the free energy diagram displays a HP-like cusp for low $\tilde\Phi$ but becomes smooth for $\tilde\Phi>\Phi_c$ (see Fig.~\ref{fig:chargedbulk} in App.~\ref{appA} for the bulk version of the free energy diagram). However, while the deconfined phase here has $W>0$ for $\tilde\Omega  R\geq 1$, that of the charged case has $W \leq 0$ for $\tilde \Phi \geq \Phi_c$. This explains the difference between the co-existence diagrams in the rotating and charged case: in the former, the phase that dominates the grand canonical ensemble for   $\tilde \Omega R\geq 1$ is the confined phase, while in the latter the phase that dominates for $\tilde \Phi \geq \Phi_c$ is   the deconfined phase.

\begin{figure}
    \centering
    \includegraphics[scale=0.8]{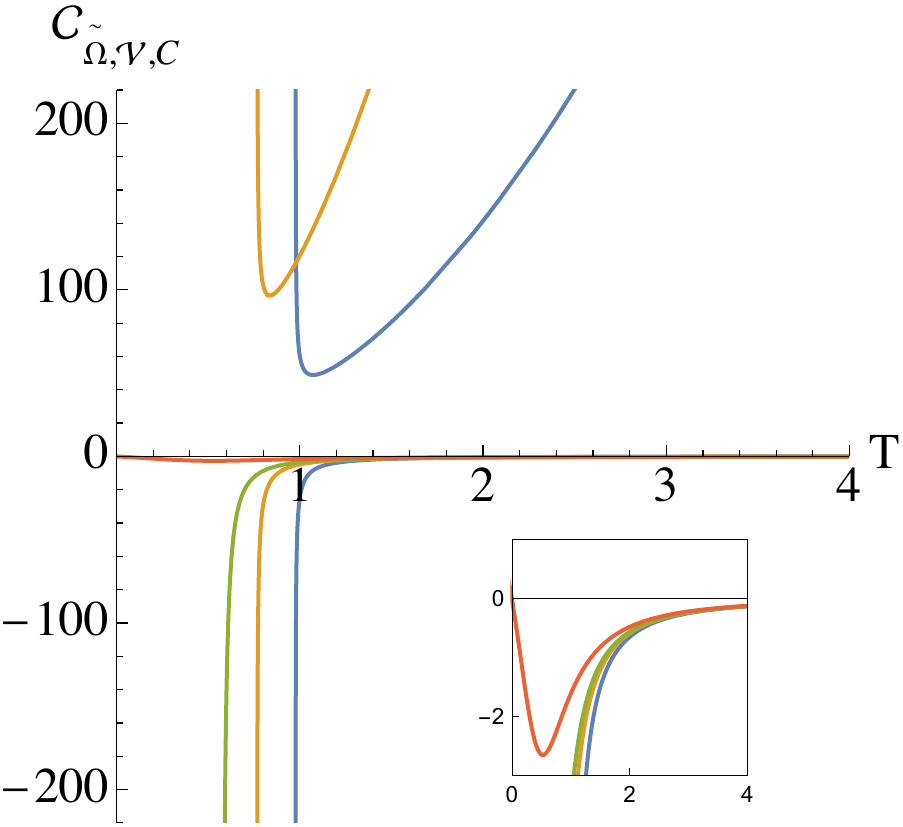}
    \includegraphics[scale=0.8]{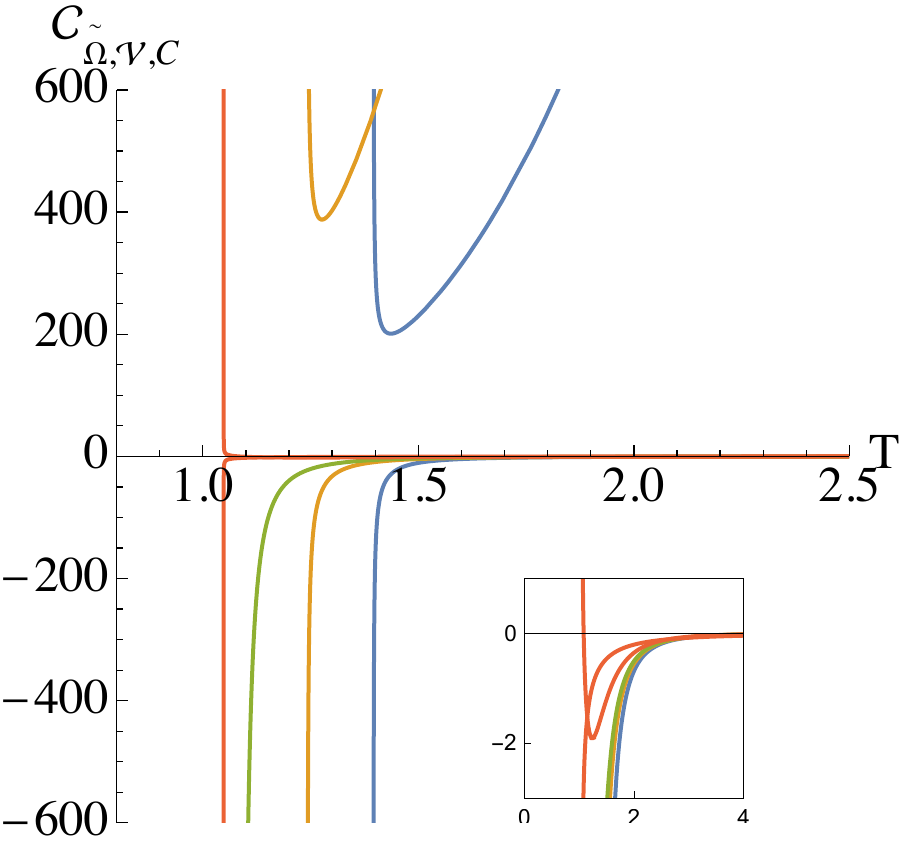}
    \caption{Heat capacity $\mathcal{C}_{\tilde \Omega,\vol,C}$ against temperature $T$ for  $\vol=1=C$, $d=4$ (left), and $d=6$ (right). The curves correspond to $\tilde\Omega R=1/20$ (blue), $\tilde\Omega R=5/6$ (yellow), $\tilde\Omega R=1$ (green) and $\tilde\Omega R=3/2$ (red), the same values as in Fig.~\ref{fig:WT}. For each $\tilde\Omega$, the position of the vertical asymptote (if one exists) happens at the value of $T$ at which the $W-T$ diagram experiences a cusp. For $\tilde\Omega R<1 $, the heat capacity has two branches: the positive branch corresponds to the  lower branch   in Fig.~\ref{fig:WT}, while the negative  $\mathcal{C}_{\tilde \Omega,\vol,C}$ branch corresponds to the upper branch. For $\tilde \Omega R = 1$, $\mathcal{C}_{\tilde \Omega,\vol,C}<0$. For $\tilde \Omega R > 1$, $\mathcal{C}_{\tilde \Omega,\vol,C}<0$ in $d=4$, while it has two branches in $d=6$ (see insets for zoom-in around origin). The (red) branch with a minimum point corresponds to the  lower (black) branch in Fig.~\ref{fig:WT}. }
    \label{fig:HComega}
\end{figure}

\subsubsection{Heat capacity and thermal stability}

We now examine the stability of the different phases as indicated by their heat capacity,
\begin{align}
    \mathcal{C}_{\tilde{\Omega} ,\vol,C}
   &=-\left(\frac{\sqrt{z-\tilde{\Omega} R z^2}}{\sqrt{\tilde{\Omega} R-z}}\right)^d \frac{4 \pi  C \left(\tilde{\Omega} R-z\right)}{z \left(\tilde{\Omega} R z-1\right)^2}\,\,\times
\\
  &\qquad \times \, 
  \frac{ \left(-\left((d-4) z^2\right)+2 (d-3) \tilde{\Omega} R z-d+2\right) \left((d-3) \tilde{\Omega} R
   \left(z^2-1\right)+2 \tilde{\Omega}^2R^2 z-2 z\right)}{(d-3) \tilde{\Omega} R \left(z^4+6 z^2+1\right)-2 \tilde{\Omega}^2R^2 z \left((d-2)
   z^2+d-4\right)-2 z \left((d-4) z^2+d-2\right)}\,. \nonumber 
\end{align}
%where introduced the notation $\tilde \Omega R =: \hat{\Omega}$.
We plot the heat capacity $\mathcal{C}_{\hat{\Omega},\vol,C}$ against the temperature~ $T$   in Fig.~\ref{fig:HComega}. From this  we see that for $\tilde\Omega R<1 $ the   lower branch of the $W-T$ diagram   has positive heat capacity and is therefore thermodynamically stable, while the upper branch has negative heat capacity. Moreover, the heat capacity is negative in $d=4$ for $\tilde \Omega R \ge 1$, and in $d=6$ it is negative for $\tilde \Omega R=1$. Further, the heat capacity of the  solution with $\tilde \Omega R \ge 1$ in $d=6$ has an interesting feature: the upper branch in the $W-T$ diagram (black curve in Fig.~\ref{fig:WT}) has negative heat capacity, while   the lower branch has negative heat capacity for large temperatures but positive heat capacity for small temperatures. The positive heat capacity  for small temperatures is a novel feature in six dimensions compared to four dimensions. 

%\begin{figure}
%    \centering
%    \includegraphics[scale=0.75]{figures/phiVCWT.pdf}
 %   \hspace{1cm}
%        \includegraphics[scale=0.75]{figures/phiTPD.pdf}
%    \caption{Fixed potential ensemble for charged AdS black hole. \textbf{Left}: $W - T$ plot for $\tilde\Phi < \Phi_c$ (blue), $\tilde\Phi=\Phi_c $ (orange) and $\tilde\Phi > \Phi_c$ (green).  \textbf{Right}: $\tilde \Phi-T$ phase diagram. \wan{do we want to include these, or just refer to the charged paper?}}
%    \label{fig:charged grand}
%\end{figure}

\begin{figure}
    \centering
    \includegraphics[scale=0.7]{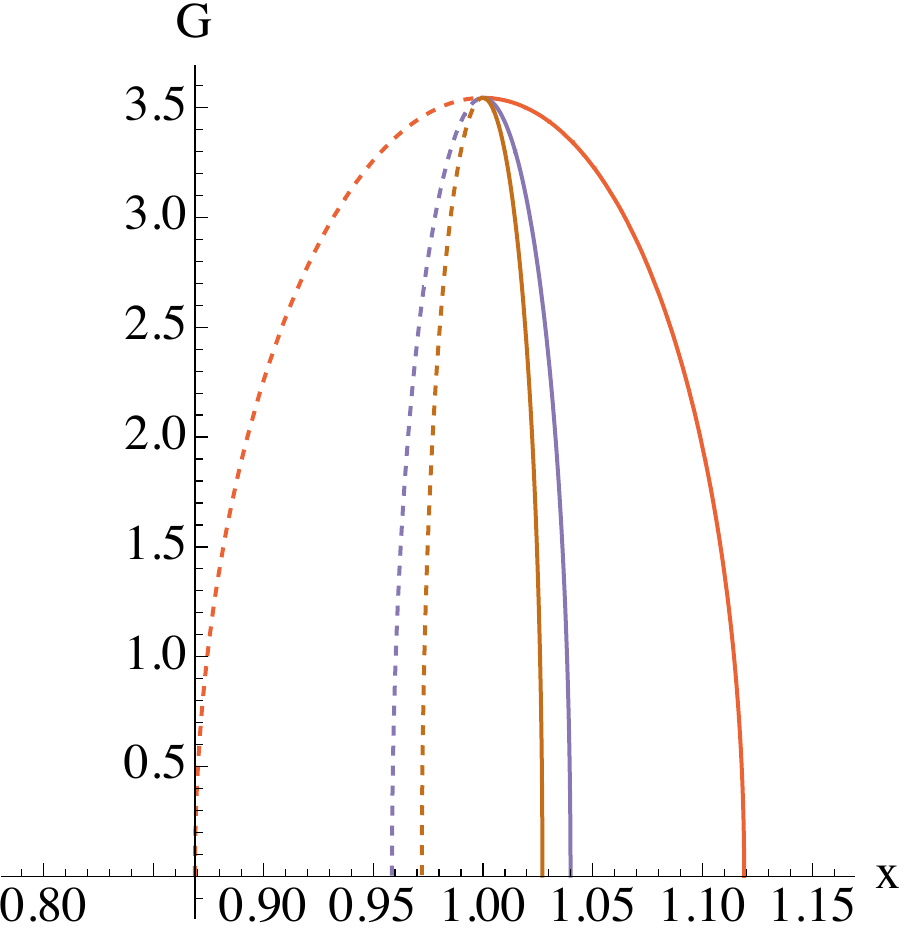}
    \includegraphics[scale=0.7]{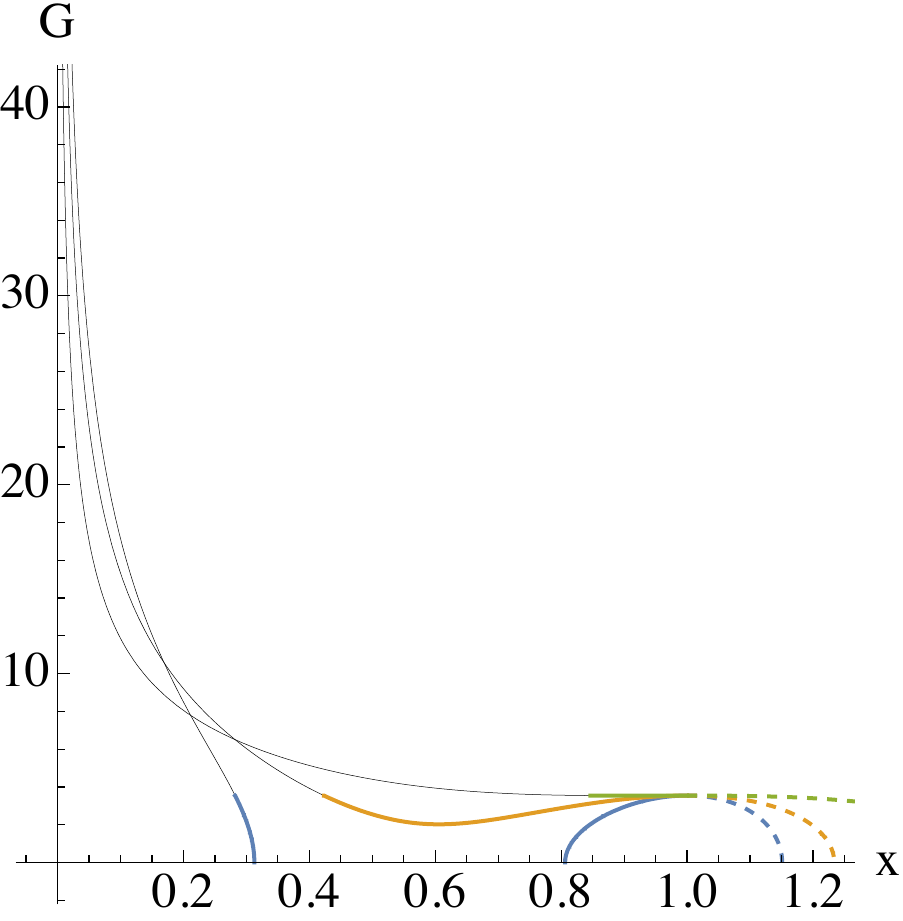}
    \includegraphics[scale=0.7]{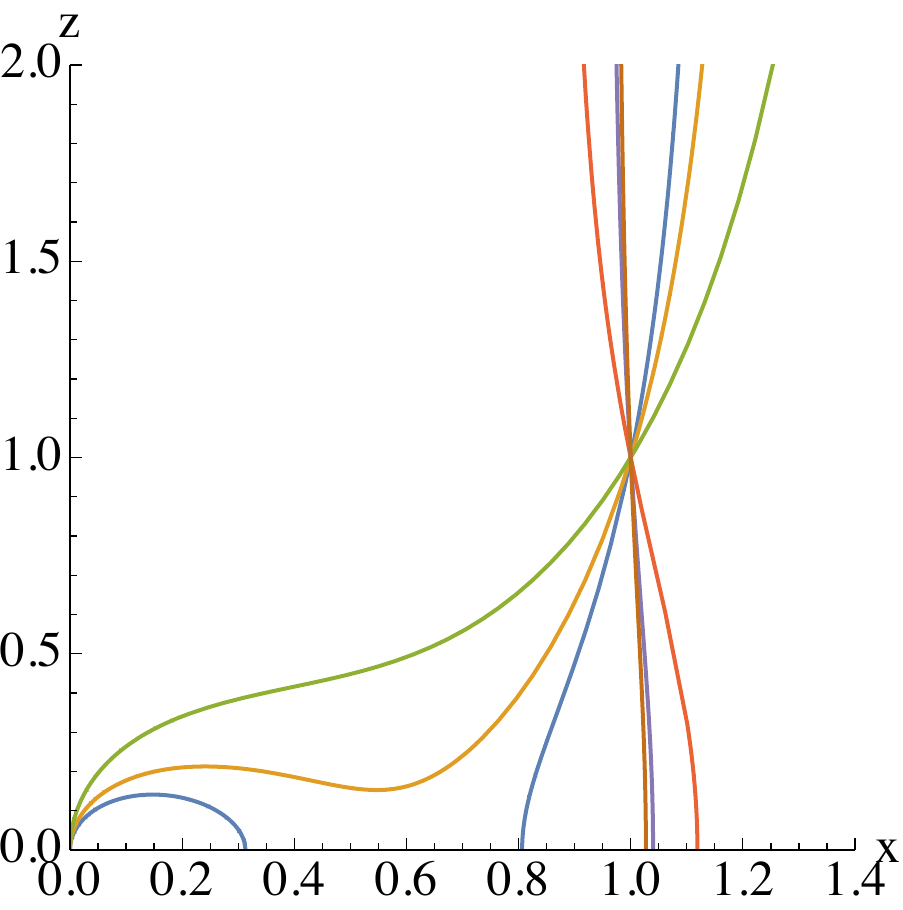}
    \caption{Physical ranges of $x$. The parameters used for the plots here are $d=4$, $\vol=1$, $J=1$, $\mu=-1$ (red), $\mu = -0.3$ (purple), $\mu = -0.2$ (brown), $\mu=1$ (blue), $\mu=3/2$ (yellow), $\mu=3$ (green). The physical range of $x$ is limited to   those values for which $z\leq 1$. From the bottom figure, observe that for $\mu>0$, this corresponds to $x<1$ while for $\mu<0$, this corresponds to $x>1$. The physical ranges are depicted using solid lines in the $G-x$ plots, while the $z>1$ regions are dashed. The same applies to the other plots in this section. The black lines correspond to superradiant states. }
    \label{fig:GxTx4D}
\end{figure}

\subsection{Novel ensemble: $G(T,J,\mathcal V,\mu)$}
\label{novelensemble}
%To study this ensemble, we write $z = z(x,R,\mu)$ using~\eqref{mu} and $C=C(J,z(x,R,\mu),x)$ using~\eqref{J}. Substituting these into the expression~\eqref{T} and the above expression for $G$ allows us to study the free-energy behaviour at fixed $(J,V,\mu)$. 

Finally, we study the ensemble in which the chemical potential $\mu$ for the central charge is kept fixed, while the central charge itself is allowed to vary. Although the physical meaning of this ensemble is not entirely clear, fixing $\mu$ corresponds to fixing $W/C$, or in other words fixing the thermal free energy per degree of freedom.

%\subsubsection{Free Energy}
The free energy $G$ and temperature $T$ in the fixed $(J,\mathcal V,\mu)$ ensemble can be expressed as
\begin{equation}
\label{G}
    G=\frac{2  C x^{d-5} z^2 \left(x^2+1\right)^2 }{R \left(z^2-1\right)^2} =J \frac{z}{R } \frac{x^2 + 1}{x^2 + z^2} =\frac{J   }{R} \sqrt{\left(1+ \frac{1-x^2}{\mu  R x^{5-d}}\right) \left(1+ \frac{x^4-x^2}{\mu  R x^{5-d}}\right)}\,,
\end{equation}
%or as
%\begin{equation}
%    G = J \frac{z}{R } \frac{x^2 + 1}{x^2 + z^2} =\frac{J   }{R} \sqrt{\left(1+ \frac{1-x^2}{\mu  R x^{5-d}}\right) \left(1+ \frac{x^4-x^2}{\mu  R x^{5-d}}\right)} 
    %=  \frac{J /R }{\mu R x^{5-d} } \sqrt{\left(1-x^2+\mu  R x^{5-d}\right) \left(x^4-x^2+\mu  R x^{5-d}\right)} 
    %=% \sqrt{\frac{\mu  R x^{5-d}-x^2 (1-x^2 ) }{\mu  R x^{5-d}+1-x^2}}\frac{  \left(\mu  R x^{5-d}+1-x^2\right)}{\mu  R}
%\end{equation}
and 
\begin{equation}
\label{T3}
    T = \frac{1}{4 \pi  \mu  R^2  } \left [ \mu  R   \left((d-1) x+\frac{d-5}{x}\right)-2 x^{d-4}\left(x^2-1\right)
   \right]\,,
\end{equation}
by using the following expression for the rotation parameter $z$:
\begin{equation}
\label{z}
    z = \sqrt{\frac{x^4 -x^2 +\mu  R x^{5-d} }{1-x^2+\mu  R x^{5-d}}}\,.
\end{equation}
The above expressions for $G(x,J,\vol,\mu)$ and $T(x,\vol,\mu)$ allow us to parameterically plot the $G-T$ diagram using $x$ as the parameter. However, the physical range of $x$ is limited to values  for which $z(x)<1$, since from Eqs.~\eqref{xz} $z>1$ corresponds to an overspinning black hole with $a>L$, while Eq.~\eqref{entropy} shows that $S<0$ when $z>1$. 
Using~\eqref{z} we plot $z(x)$ in the bottom diagram of Fig. \ref{fig:GxTx4D} for $d=4$ as an illustration. From this and~\eqref{z}, we see that the physical ranges of $x$ are $x<1$ for $\mu>0$ and $x>1$ for $\mu<0$. Nonetheless, we continue to plot the unphysical ranges
%\footnote{In \cite{where?}, the authors showed that the transformation $z\rightarrow -z$ \wan[Is this the right transformation?] leads to another regular rotating black hole. However, the transformed black holes are in completely different thermodynamic states as the original i.e., they have different $J$ and $\mu$, and hence are irrelevant for the following thermodynamic considerations.}
in the following $G-T$ figures, but denoting the $S<0$ region with \textit{dashed lines}. As in the previous two ensembles, we shall study the $G-T$ behaviour in $d=4$ and $d=6$.

\begin{figure}
    \centering
    \includegraphics[scale=0.7]{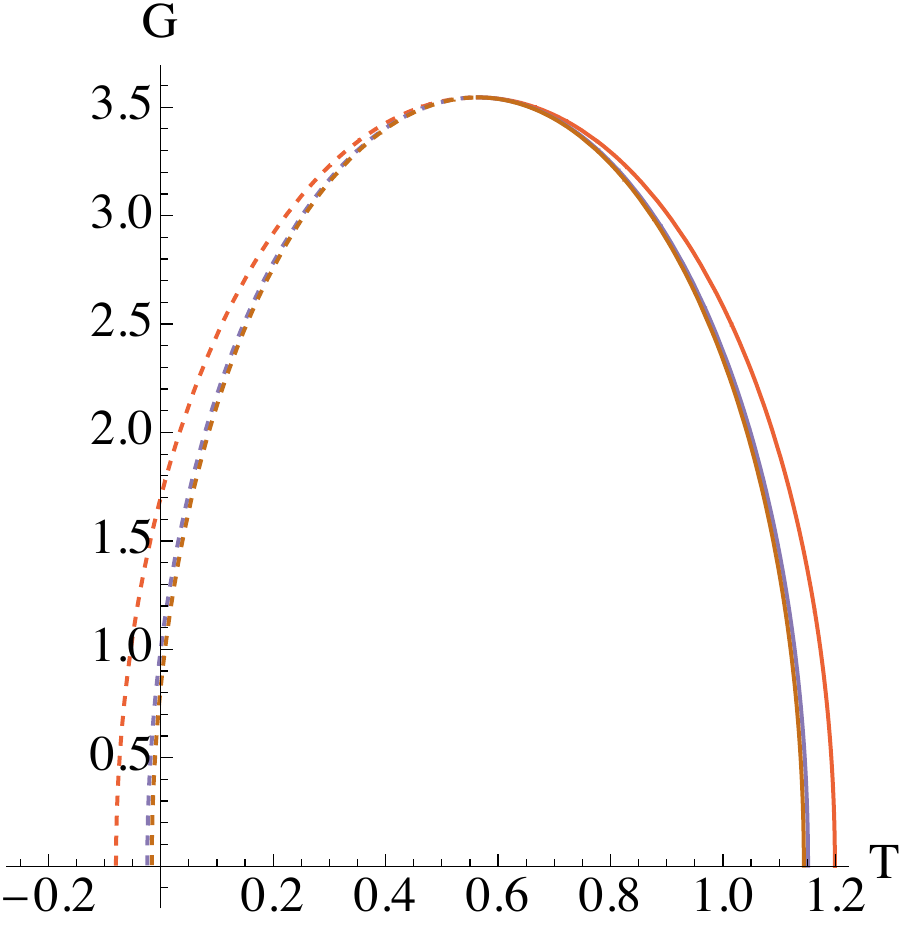}
      \includegraphics[scale=0.7]{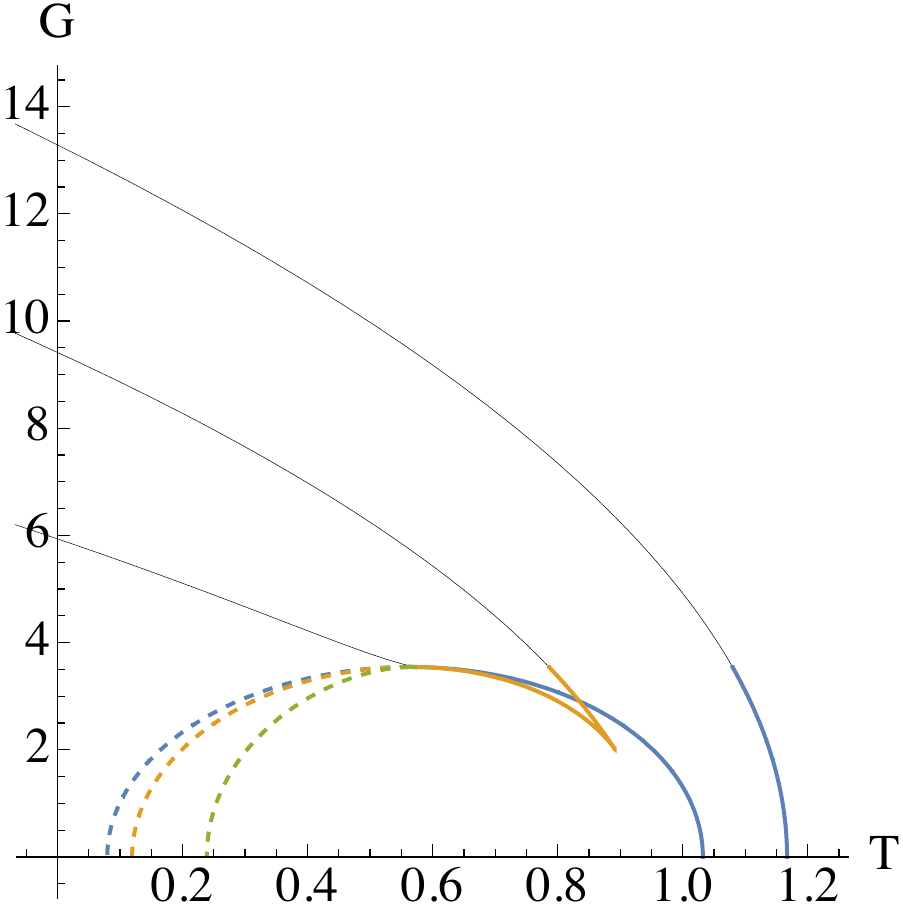}
\includegraphics[scale=0.45]{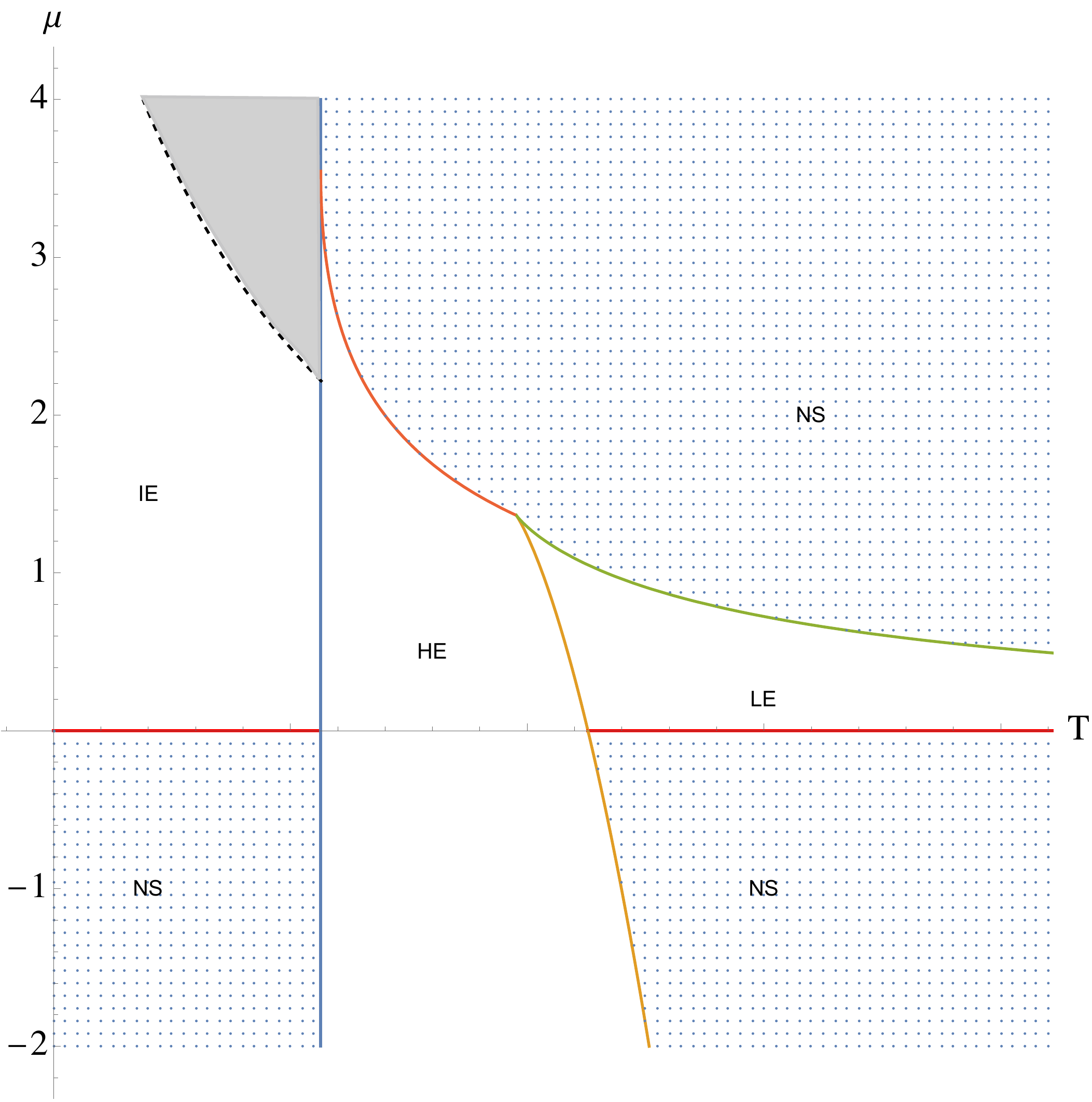}
    \caption{Thermodynamic behaviour in fixed $(\mathcal{J},\mathcal{V},\mu)$ ensemble for $d=4$. The figures here are plotted for $\vol=1$, $J=1$. \textbf{Top left:} $G-T$ free energy diagram for $\mu<0$,  $\mu=-1$ (red), $\mu = -0.3$ (purple), $\mu = -0.2$ (brown).  \textbf{Top right:}   $\mu>0$,  $\mu=1$ (blue), $\mu=3/2$ (yellow), $\mu=3$ (green); {the black portions of the curves denote the solutions with $\tilde \Omega R > 1$, where superradiant instabilities are present in the bulk. }\textbf{Bottom:} Co-existence phase diagram. Across each solid curve  a zeroth-order phase transition takes place between the indicated phases. The IE phase is further split into a region with positive heat capacity $\mathcal{C}_{\mathcal{J},\mathcal{V},\mu}$ and a region with negative heat capacity (shaded).
     }
    \label{fig:muT4d}
\end{figure}

\subsubsection{$d=4$: Zeroth-order phase transitions}

The first feature of this ensemble is the different behaviour exhibited by the system when $\mu<0$ and $\mu>0$. For example in $d=4$, the $G-T$ diagram shows a single ``dome'' (see Fig.~\ref{fig:muT4d}, top left) for $\mu<0$, while for small $\mu>0$, the dome is accompanied by a bigger quarter-arc (e.g. blue curve in Fig.~\ref{fig:muT4d}, top right). In both cases, the $G-T$ graphs for different $\mu$ all intersect at the dashed-solid boundary point, $(T_i,G_i)=((d-3)/(2 \pi R),J/R)$, corresponding to $x=1$. 

Hence, the system has only a single phase when $\mu<0$ but we can identify multiple phases for $\mu>0$. As in the above ensembles, we label these phases according to their relative $x$ values. For $\mu=1$ (blue) in Fig.~\ref{fig:muT4d}, $x$ is smaller on the upper quarter-arc (cf. Fig. \ref{fig:GxTx4D}, bottom), which we label the low-entropy (LE) phase, whereas the high-entrop (HE) phase refers to    the inner dome. As~$\mu$ gets larger, 
the domes move closer together, eventually joining at a cusp for sufficiently large~$\mu$. For larger~$\mu$ the cusp moves upward and leftward, as shown by the yellow curve.

\begin{figure}
    \centering
      \includegraphics[scale = 0.8]{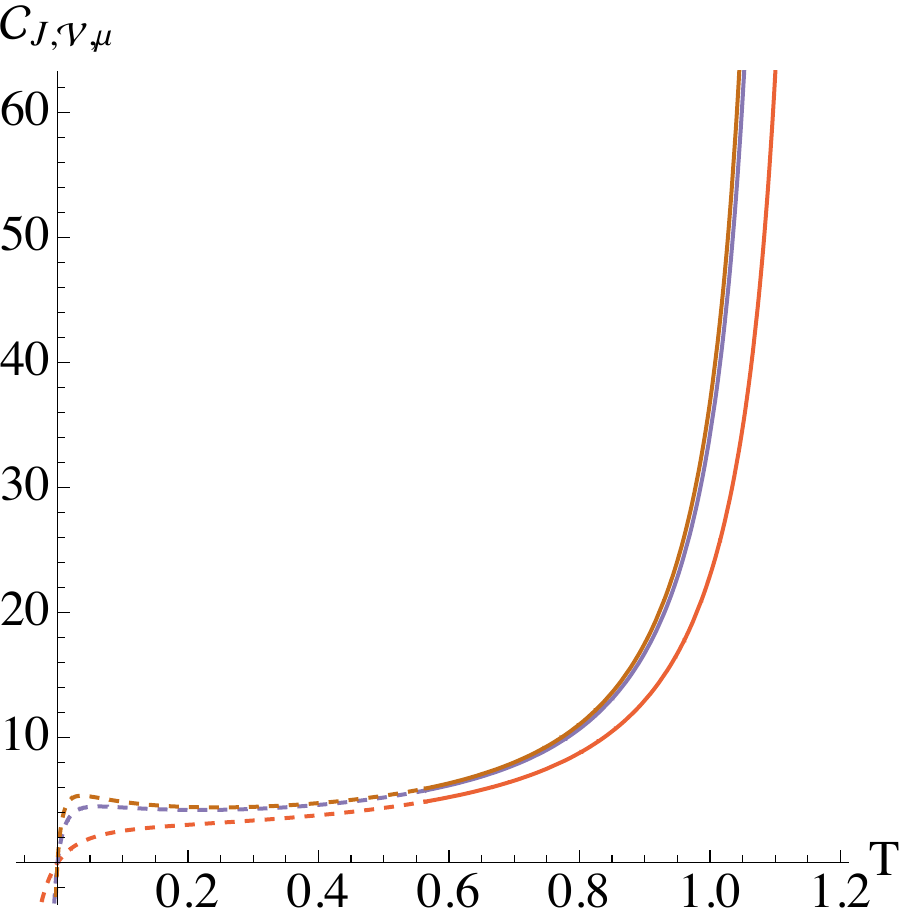}
    \includegraphics[scale=0.8]{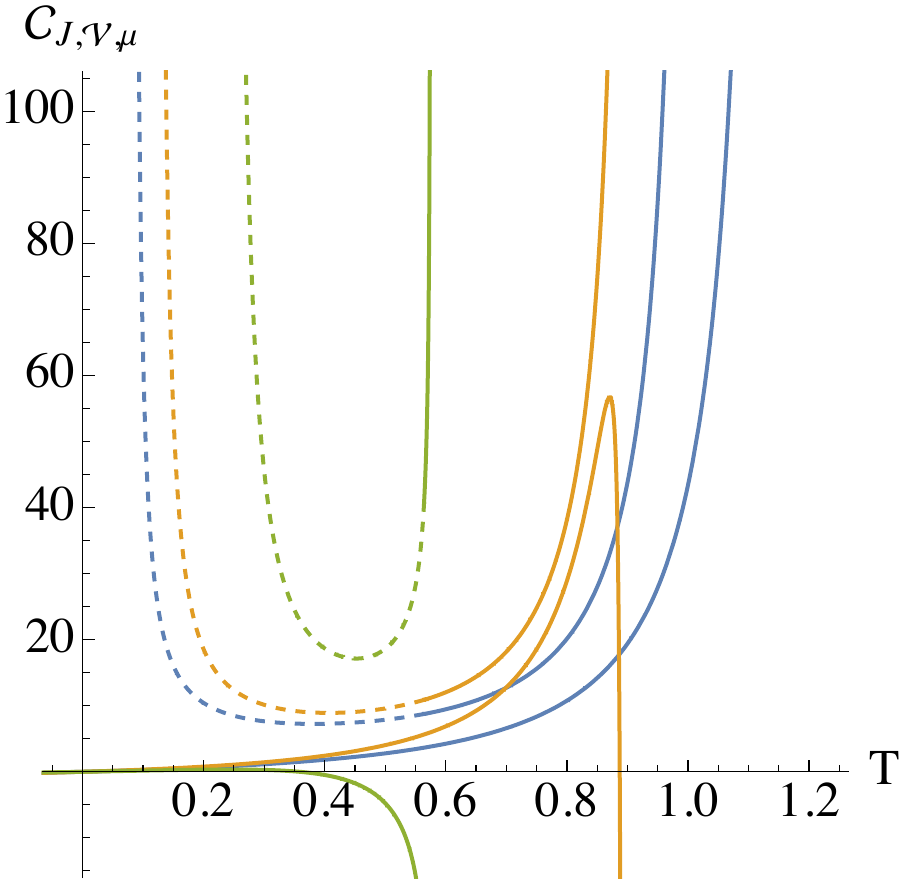}
  \caption{Heat capacities in fixed $(J,\mathcal{V},\mu)$ ensemble for $d=4$. \textbf{Left:} $\mu<0$,  $\mu=-1$ (red), $\mu = -0.3$ (purple), $\mu = -0.2$ (brown); \textbf{ Right:}   $\mu>0$,  $\mu=1$ (blue), $\mu=3/2$ (yellow), $\mu=3$ (green). }
    \label{fig:GTHC4d}
\end{figure}

The second feature to note in this ensemble is that all phase transitions are accompanied with a jump in $G$; in other words, they are \emph{zeroth-order phase transitions}.   As usual, for any fixed $\mu$ if the $G-T$ diagram is multi-branched (considering only the solid  $z>1$  regions), the branch with the lowest free energy~$G$ is thermodynamically favoured. A phase transition between different branches is implied whenever the $G-$minimizing branch changes.  The bottom diagram of Fig.~\ref{fig:muT4d} shows all the implied phase transitions on the $\mu-T$ phase diagram. For example at $\mu=1$, corresponding to the blue curve at the top right, there is no solution (NS) at high temperature. As $T$ decreases, the LE 
solution is admitted; the
system crosses the NS/LE boundary on the $\mu-T$ diagram. As $T$ decreases further, the HE phase 
emerges with a lower value of~$G$, giving rise to a LE/HE zeroth-order phase transition. The HE phase terminates at the $z=x=1$ boundary (the solid-dashed boundary point of the $G-T$ curve) and the system undergoes another zeroth-order phase transition to the intermediate entropy (IE) phase. The IE phase is further divided into a stable region with positive heat capacity $\mathcal{C}_{\mathcal{J},\mathcal{V},\mu}$, and an unstable region (shaded) with  negative heat capacity. The other phases in the $\mu-T$ plot are all stable (cf.~Fig.~\ref{fig:GTHC4d}).

The explicit expression for the heat capacity in $d=4$ is
\begin{align}
\mathcal{C}^{d=4}_{\mathcal{J},\mathcal{V},\mu} & = \frac{\pi  \mathcal{J} \mu  R \sqrt{x} \left(4 \mu  R x^3+\left(x^2-1\right)^3\right) \left(x \left(3 \mu  R x-2 x^2+2\right)-\mu 
   R\right)}{\left(-\mu  R-x^3+x\right)^{3/2} \left(-\mu  R x+x^2-1\right)^{3/2} \left(\mu  R+3 \mu  R x^2-4 x^3\right)}\,,\
\end{align}
which we plot in Fig.~\ref{fig:GTHC4d}.

\subsubsection{$d=6$: Unstable small entropy phase}

\begin{figure}
    \centering
    \includegraphics[trim= 4.3cm 1cm 8cm 0cm, clip=true, scale=0.27]{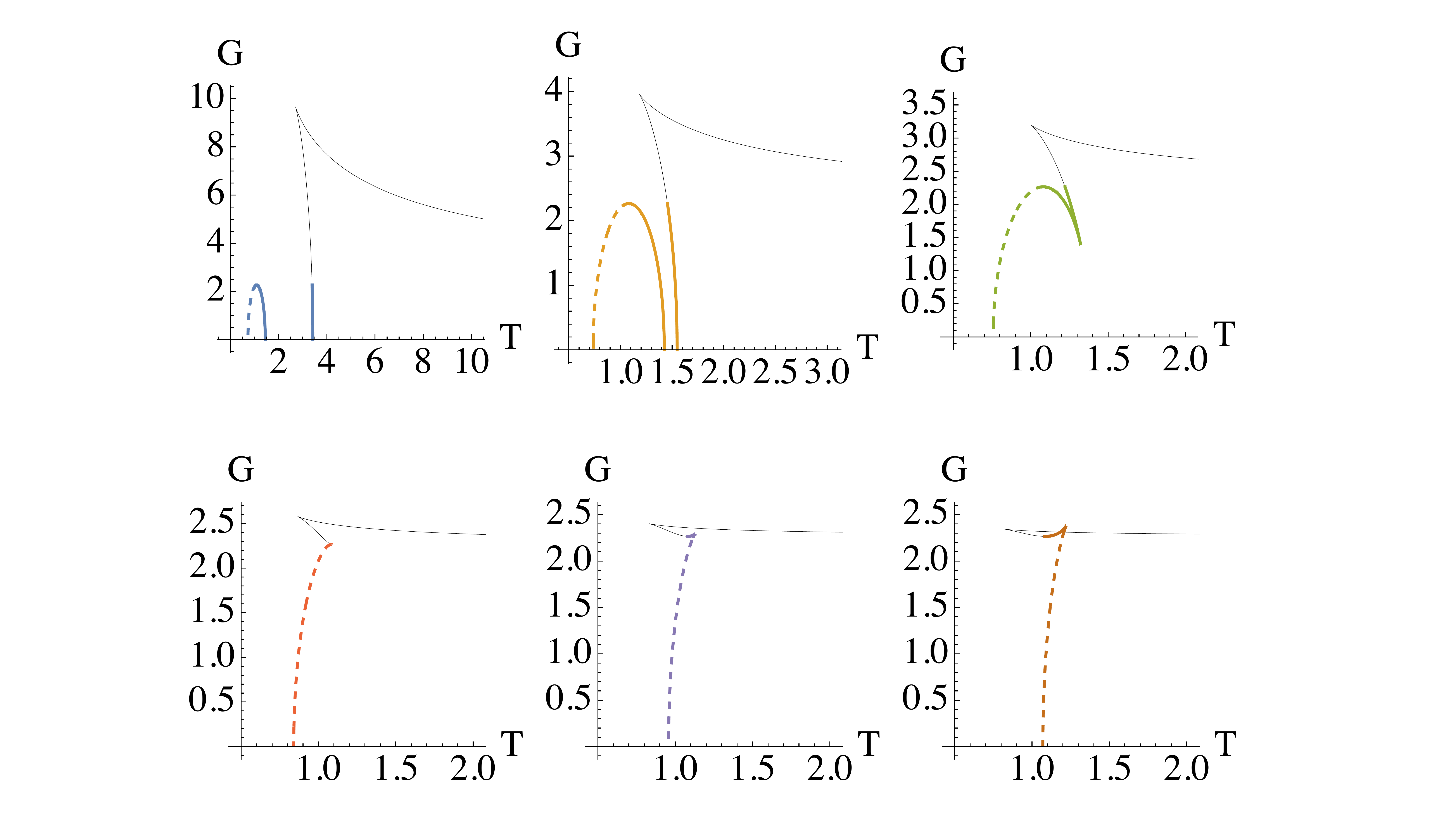}
    \includegraphics[trim= 4cm 2cm 6cm 0cm, clip=true,scale=0.255]{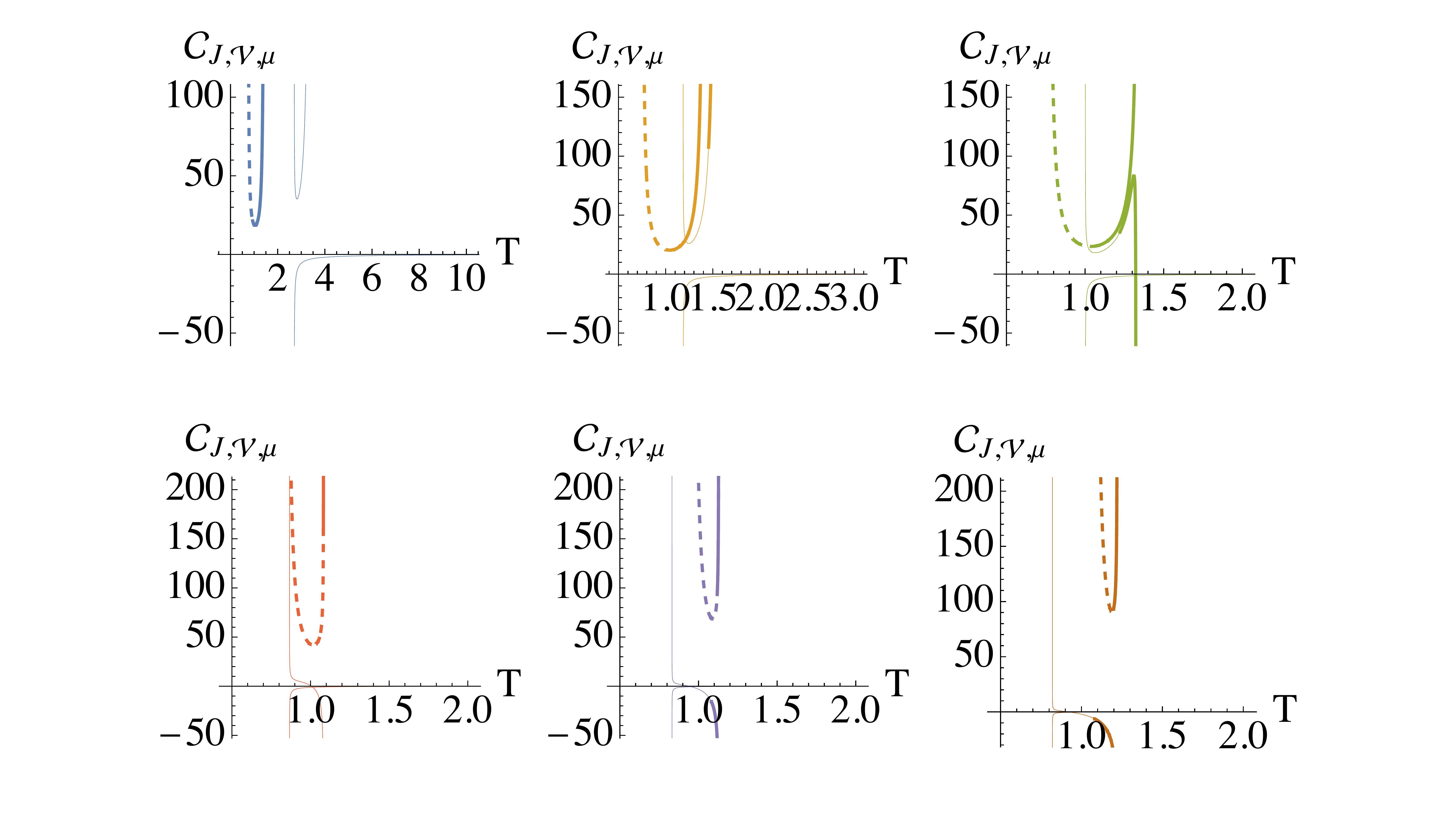}
    \caption{$G-T$ diagrams and heat capacities for $\mu>0$ in $d=6$. These are plotted for different illustrative $\mu$ values. Specifically, the parameters used here are $\vol=J=1$, and for the $G-T$ digrams on the first two rows, starting from the top left, we plotted $\mu=0.01, 0.2, 0.5,2,5,9$\,.  The black portions of the curves denote the solutions with 
    $\tilde \Omega R > 1$.  
    The same parameters are used for the analogous heat capacity plots in the last two rows.
    % [comment on IE regions; positive for $\mu<0$]
    }
    \label{fig:GT6dp}
\end{figure}

\begin{figure}
    \centering
    \includegraphics[scale=0.7]{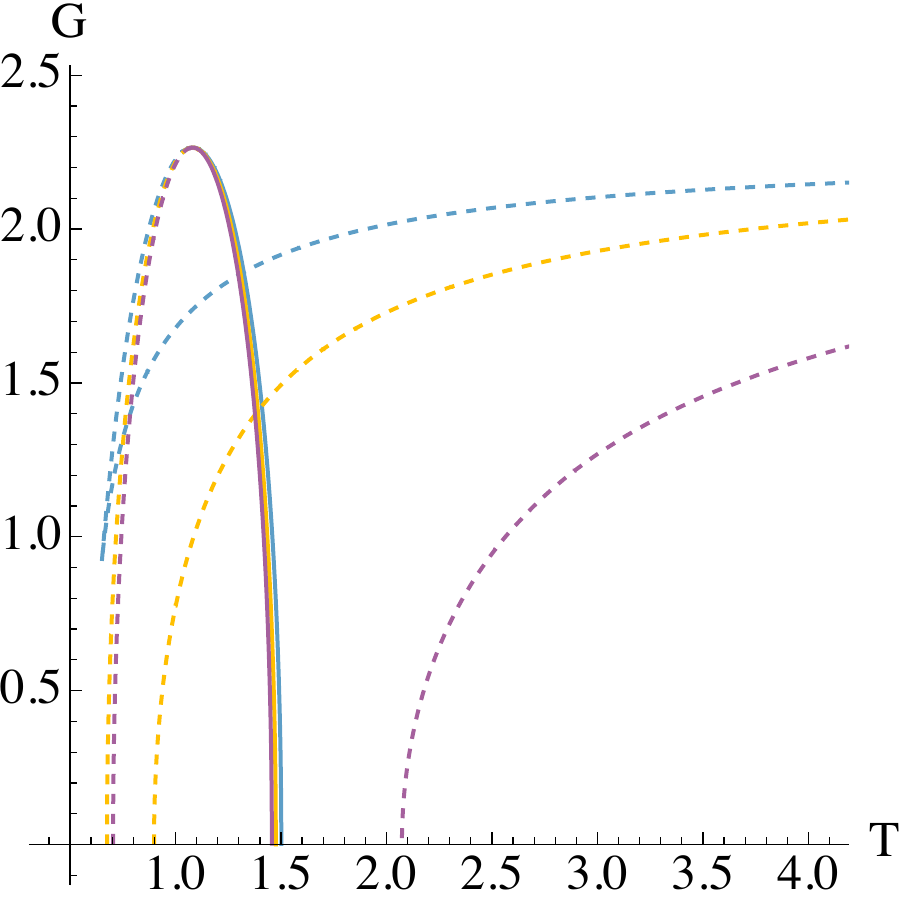}
    \includegraphics[scale=0.7]{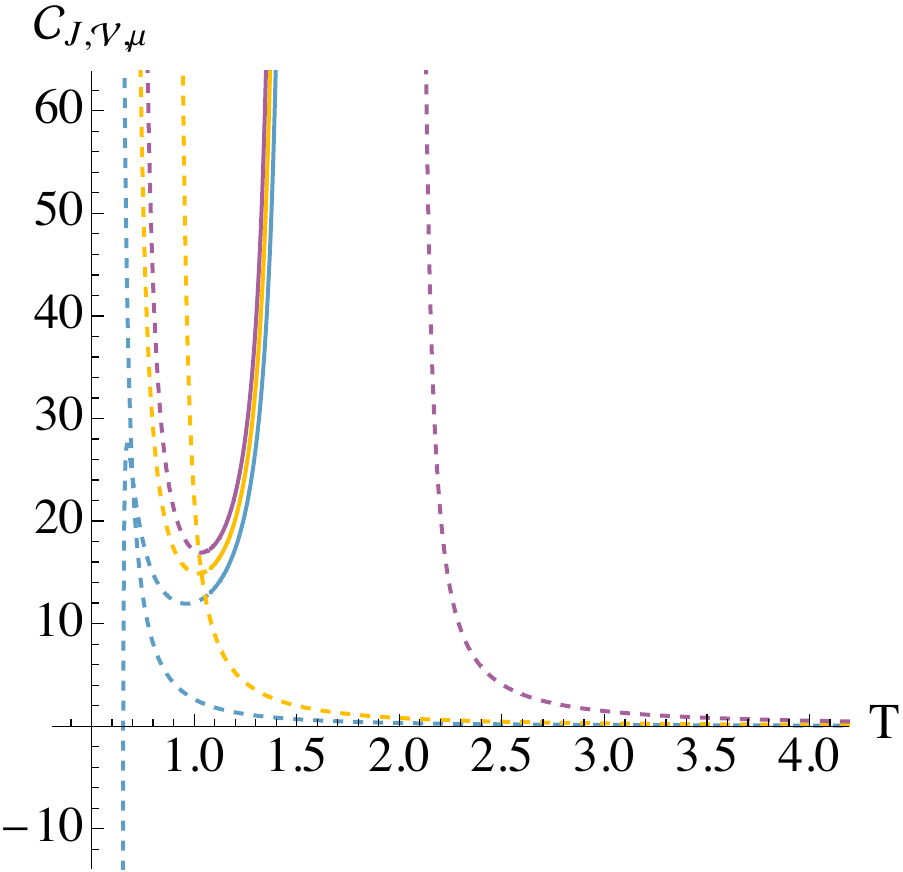}
    \caption{$G-T$ diagram and heat capacities for fixed $(J, \vol, \mu)$ ensemble  with $\mu<0$ in $d=6$. The parameters used here are $\vol=J=1$, $\mu=-1$ (blue), $\mu=-0.5$ (yellow), $\mu=-0.2$ (purple).}
    \label{fig:GT6dn}
\end{figure}

\begin{figure}
    \centering
    \includegraphics[scale=1]{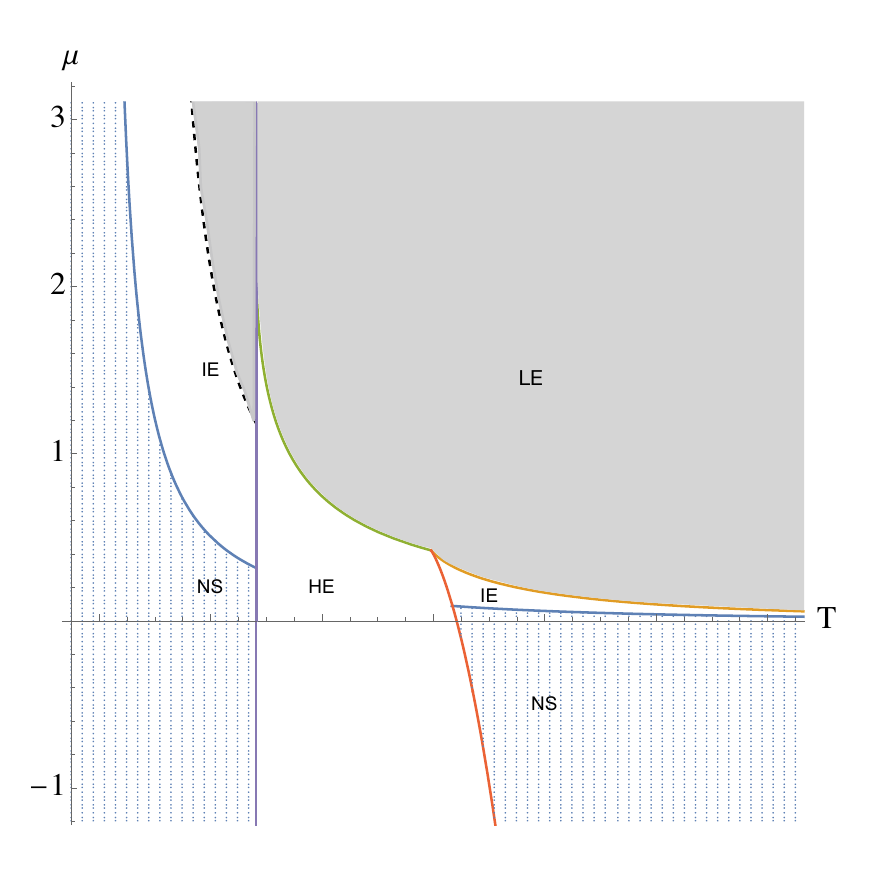}
    \caption{$\mu-T$ phase diagram in $d=6$, which follows from analysing the $G-T$ free energy diagrams in Fig. \ref{fig:GT6dp} and \ref{fig:GT6dn}. Across each curve  a zeroth-order phase transition (finite jump in $G$) takes place between the labeled phases. There are also regions (NS) where no solution exists at the given $(\mu, T, \mathcal{V}=1)$. The left IE phase is further split into a region with positive heat capacity $\mathcal{C}_{\mathcal{J},\mathcal{V},\mu}$ and a region with negative heat capacity (shaded). % \MR{LE phase given a grey color}
    }
    \label{fig: GT6DphaseD}
\end{figure}

A similar analysis can be done for $d=6$. The $G-T$ and corresponding $C_{J,\vol,\mu}-T$ diagrams for $\mu>0$ and $\mu<0$ are shown in Figs. \ref{fig:GT6dp} and \ref{fig:GT6dn}, respectively, where
\begin{align}
%   \\
%   \mathcal{C}^{d=5}_{\mathcal{J},\mathcal{V},\mu} & =-\frac{2 \mathcal{J} \mu  R x \left(-2 \mu  R+x^2-1\right) \left(\mu  R \left(3 x^2-1\right)+\left(x^2-1\right)^3\right)}{\left(-2 \mu  R+3 x^2-1\right) \left(\mu  R-x^2+1\right)^3 \left(\frac{x^4-1}{\mu  R-x^2+1}+1\right)^{3/2}}\, ,
   \mathcal{C}^{d=6}_{\mathcal{J},\mathcal{V},\mu} & = \frac{\pi  \mathcal{J} \mu  R x^2 \left(\mu  R+5 \mu  R x^2-2 x^5+2 x^3\right) \left(4 \mu  R \left(2 x^2-1\right)+3 x
   \left(x^2-1\right)^3\right)}{\left(-\mu  R+x^3-x\right)^{3/2} \left(-\mu  R-x^5+x^3\right)^{3/2} \left(-\mu  R+5 \mu  R x^2-8 x^5+4
   x^3\right)}
\end{align}
is the explicit expression for the heat capacity in $d=6$.  
While these look more complicated than in $d=4$, the main information is contained in the $\mu-T$ phase diagram, Fig. \ref{fig: GT6DphaseD}. Similar to Fig. \ref{fig:muT4d}, obtaining this figure is straightforward but tedious. We leave the details to   Appendix \ref{appB} and comment here only on the main features. In this diagram, a zeroth-order phase transition again takes place across each solid curve (except of course at the NS boundary). In fact, the structure of this diagram is somewhat similar to the $d=4$ case. However, unlike $d=4$, for $\mu>0$  solutions now extend to large $T$ thus replacing the NS region in the upper right portion of the diagram in $d=4$ by an LE phase in $d=6$. This phase has negative heat capacity and so is unstable, as is evident from the lower diagrams in Fig.~\ref{fig:GT6dp}. 
Another distinction between the two cases is that the NS boundary in $d=4$ which lies on (part of) the $T-$axis is now given by two curves in the 
$\mu>0$ region, both terminating at a finite value of $\mu$.

Finally, we note from~\eqref{T3} that $T$ does not depend on $J$, while $G$ only depends on $J$ through an overall factor  in all $d$ (see Eq.~\eqref{G}). As a result, changing the value of $J$ has a trivial effect on the figures presented in this subsection: for the $G-T$ figures, changing $J$ only stretches the curves along the $G-$axis while all zeroth-order phase transitions occur at the original $T$ values. Consequently the $\mu-T$ phase diagrams are independent of $J$, unlike in the fixed $(J,\vol,C)$ ensemble.

% \begin{figure}
%     \centering
%     \includegraphics{figures/Zvx6d.pdf}
%     \caption{$d=6$,$V=J=1$, $\mu=-1$ (blue),$\mu=-0.3$ (yellow), $\mu=-0.2$ (brown), $\mu=0.1$ (red), $\mu=0.5$ (purple)}
%     \label{fig:zx6d}
% \end{figure}

\clearpage

\section{Discussion}\label{sec4}
We have studied the thermodynamic phase transitions of thermal CFT states  dual to Kerr-AdS black holes. The inclusion of  the conjugate thermodynamic pair 
$(C,\mu)$ (the 
central charge and its associated chemical potential)  increases the number of possible thermodynamic ensembles to eight -- we have uncovered interesting phase behavior in three of them. 

In previous studies, it was argued that  
 the (inverse) central charge plays a similar role to the thermodynamic pressure $P \propto 1/L^2$ in  the bulk thermodynamics of AdS black holes  \cite{Gunasekaran12,Altamirano:2013ane,Altamirano:2014tva}.  However as explained in \cite{Visser:2021eqk,Cong:2021jgb, Ahmed:2023snm} and in the introduction,  this %the previous setting
does not lead to a satisfactory duality between the bulk and boundary thermodynamics. To achieve this latter goal, one needs to either view   Newton's constant $G_N$ as a variable whose variation is related to variations in the CFT central charge, or introduce a new scale $R$ for the radius of the CFT sphere.  We find the latter scenario much more plausible. In addition, a reshuffling and rescaling of thermodynamic variables is needed to arrive at the holographic dictionary~\eqref{extendeddictionary} which gives a clean duality between bulk and boundary first laws and Smarr relations.

In this new setting, we find that all the interesting phase behaviours in \cite{Chamblin:1999tk,Gunasekaran12,Altamirano:2013ane,Altamirano:2014tva} are preserved by the CFT -- this is unsurprising, since we are simply replacing $P\propto 1/L^2$ by the new thermodynamic variable $C \propto L^{d-2}$. In particular, there continues to be swallowtail criticality  in the fixed $(J,\mathcal V, C)$ ensemble for $d=4$, as well as reentrant phase transitions  for  $d=6$. However, we also note some new features. The $1/C-T$ coexistence curve is negatively sloped and cuts the $T-$axis in $d=4$ at the Hawking-Page temperature $T_{\text{HP}}=1/\pi R$. Furthermore, the critical temperature is independent of $C$ and $J$, unlike in  \cite{Gunasekaran12,Altamirano:2013ane,Altamirano:2014tva}, where the critical temperature was found to be $\propto 1/\sqrt{J}$. This difference can also be easily understood as follows: our CFT temperature has been rescaled as compared to the bulk Hawking temperature $T_{H}$ in Eq.~\eqref{extendeddictionary}, 
specifically, we have $T\propto T_{H}/\sqrt{P}$. At the critical point, the old bulk pressure behaves as $P_{crit}\propto 1/J$, thus cancelling the $1/\sqrt{J}$ dependence of the critical temperature. %: $T_{crit}\propto T_{H,crit}/\sqrt{\tilde P_{crit}}\propto 1/\sqrt{J} \cdot \sqrt{J} = 1$. 

We emphasise that the role of the central charge $C$ here is unconventional insofar as studies of phase transitions of a given system generally relate to the near equilibrium dynamics of a fixed theory; changing the central charge $C$ relates instead to variations within an ensemble of theories\footnote{In the bulk, there are theories in which the cosmological constant $\Lambda$ can be treated as a dynamical variable \cite{Henneaux:1984ji,Teitelboim:1985dp} but an analogous method that makes $C$ dynamical has, to our knowledge, not been formulated.}.
We note that this is not without precedent. For example in \cite{DeLange:2018wbz}, the authors considered a family of two-dimensional CFTs   by taking symmetric products and studied the grand canonical ensemble by introducing a chemical potential conjugate to the number of copies of CFTs (the central charge of the symmetric product CFTs is proportional to this number). %the central charge and gave a rigorous derivation of the Cardy formula by studying the properties of the chemical potential conjugate to $C$.
In the current context, the observed swallowtail criticality is interesting in its own right --- it implies that within the current ensemble of rotating CFT theories, only those with a sufficiently large $C$ can undergo a first-order Van de Waals' like phase transition.

Next, in the fixed $(\tilde \Omega, \mathcal{V}, C)$ ensemble  we find a (de)confinement phase transition for $\tilde\Omega R<1$. This is dual to the black hole/radiation Hawking-Page like phase transition of the  Kerr-AdS black hole. 
Furthermore, in the peculiar fixed $(J, \vol, \mu)$ ensemble, we observe several novel zeroth-order phase transitions between the various CFT phases (see Figs. \ref{fig:muT4d} and \ref{fig: GT6DphaseD}). Moreover, at certain parameter ranges, the phase diagram suggests a transition from a phase with positive heat capacity to one with negative heat capacity. A similar transition was also observed in CFTs dual to charged AdS black holes in the fixed chemical potential ensemble \cite{Cong:2021jgb}. We caution that while these are interesting new features, the physical viability of zeroth-order phase transitions in nature is unknown. In addition, a physical interpretation of the fixed $\mu$ ensembles is elusive -- it is unclear how one can prepare a system with fixed $\mu$ while allowing the central charge to vary.

 We have also noted the presence in all ensembles of classical superradiant instabilities in the bulk, previously  observed in  \cite{Altamirano:2014tva}. It is an interesting question as to what this classical instability means for the CFT or even the bulk black hole as a thermodynamic system. Classically, a small perturbation to a superradiant black hole will lead to a decay to another non-superradiant black hole with slower rotation; hence one might imagine that a phase transition to a superradiant black hole is unphysical --- the end state will instead be replaced by the resultant non-superradiant black hole.  Note however that the superradiant instabilities were studied in the classical setting %(they were found to be a consequence of evolving the classical field equations of motion)
 (they follow from the instability of the  field equations for the class of superradiant Kerr-AdS black holes). It is an interesting question as to whether these can be related to what happens in the thermodynamic ensemble (where for example in the canonical ensemble $J$ is fixed by definition). 

We also note that whereas superradiant instabilities are classical,  thermodynamic phase transitions are semi-classical.  
Superradiant instabilities generally set in at much shorter timescales than thermodynamic ones. However  it may be that in some circumstances  the superradiant phase could be quasi-stable. We have thus included all possible phases (superradiant and non-superradiant). We leave the question of   understanding 
the implications of   superradiant phases 
for (holographic) black hole thermodynamics
for future study.\\ 

\noindent \emph{Note added}: We note that close to the completion of this project a paper \cite{Gong:2023ywu} appeared, which has an overlap with our current manuscript. They studied the first two ensembles, but did not consider the novel ensemble in Sec. \ref{novelensemble}.

\section*{Acknowledgements}
 D.K. is grateful for support from GA{\v C} 
23-07457S grant of the Czech Science Foundation, as well as would like to acknowledge the kind hospitality of the Perimeter Institute where this work was completed. M.R.V. is supported by SNF Postdoc Mobility grant P500PT-206877 ``Semi-classical thermodynamics of black holes and the information paradox''. 
This work was supported in part by the Natural Sciences and Engineering Research Council of Canada.
This research was supported in part by Perimeter Institute of Theoretical Physics. 
{Research at Perimeter Institute is supported in part by the Government of Canada through the Department of Innovation,
Science and Economic Development and by the  Province of Ontario through the Ministry of Colleges and Universities.
Perimeter Institute and the University of Waterloo are
situated on the Haldimand Tract, land that was promised
to the Haudenosaunee of the Six Nations of the Grand
River, and is within the territory of the Neutral, Anishnawbe, and Haudenosaunee peoples. 
}

\appendix
\section{Grand canonical ensemble in the bulk}
\label{appA}

\begin{figure}[b]
    \centering
    \includegraphics[trim= 4cm 4cm 0cm 0cm, clip=true, scale=0.26]{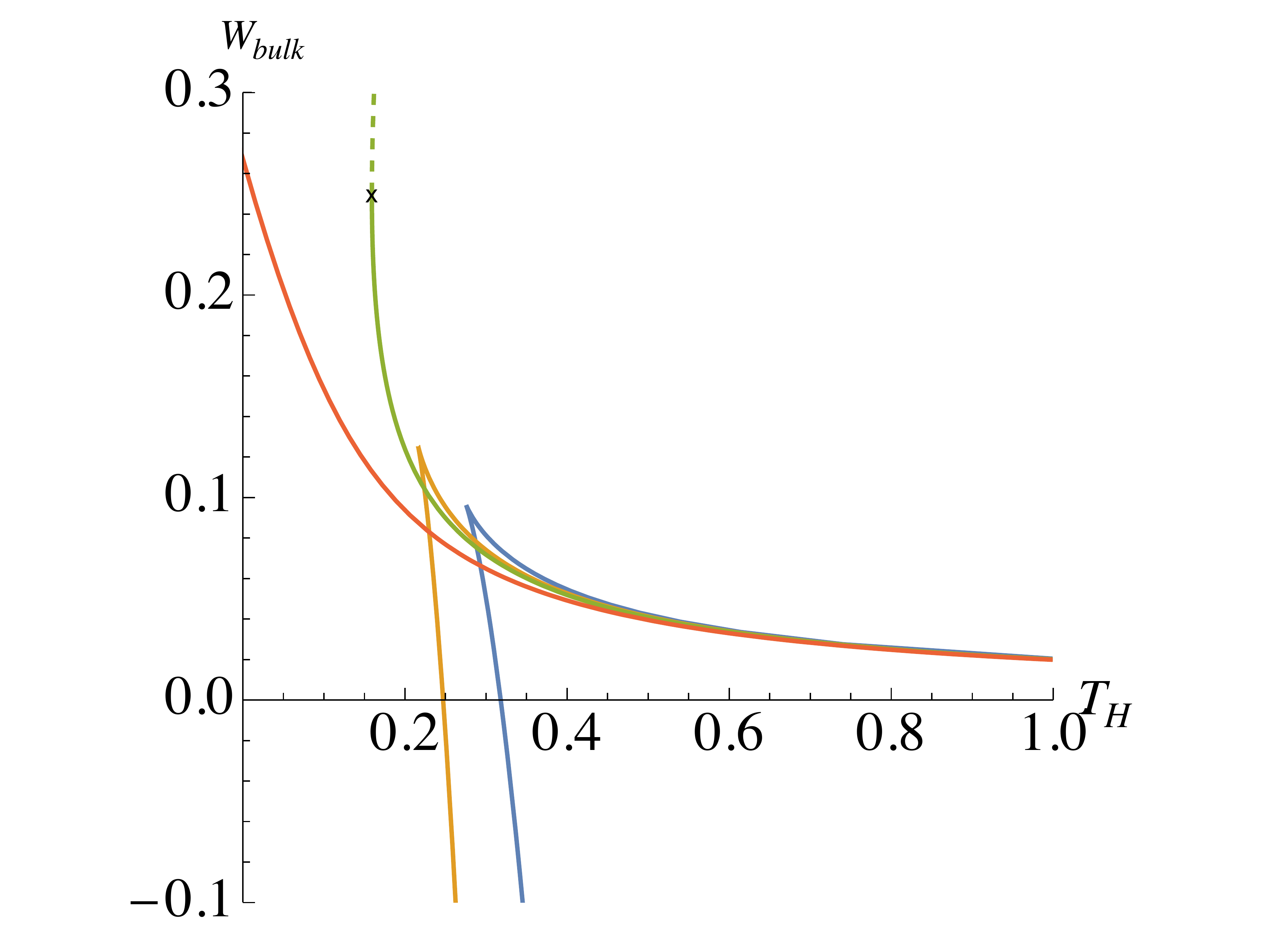}
    \includegraphics[scale=0.7]{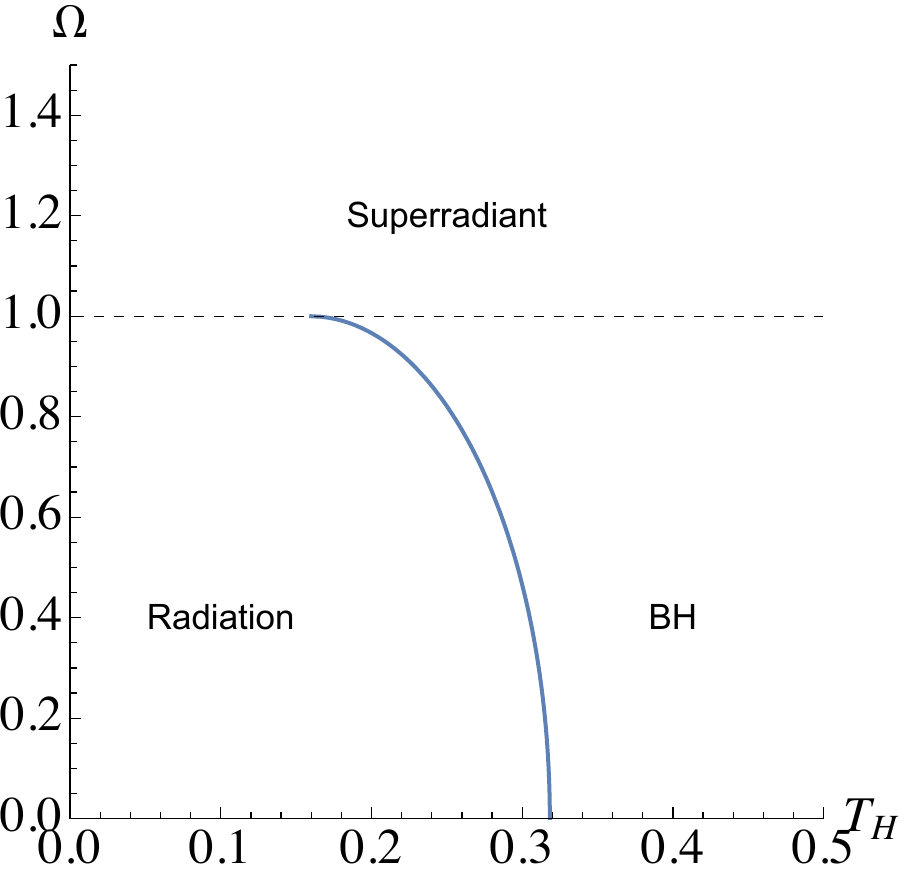}
    \caption{Bulk grand canonical ensemble (at fixed angular velocity $\Omega$) of Kerr-AdS black holes. \textbf{Left:} Free energy against temperature diagram for $\Omega L = \frac{1}{20}$ (blue), $\Omega L= \frac{5}{6}$ (yellow), $\Omega L = 1$ (green) and $\Omega L= \frac{3}{2}$ (red). The physical part of the $\Omega L=1$ curve terminates at the point ``x'', with $a = L$. \textbf{Right:} Co-existence phase diagram for $\Omega$ vs. $T_H$.}
    \label{fig:bulkgrand}
\end{figure}

\begin{figure}
    \centering
    \includegraphics[scale=0.8]{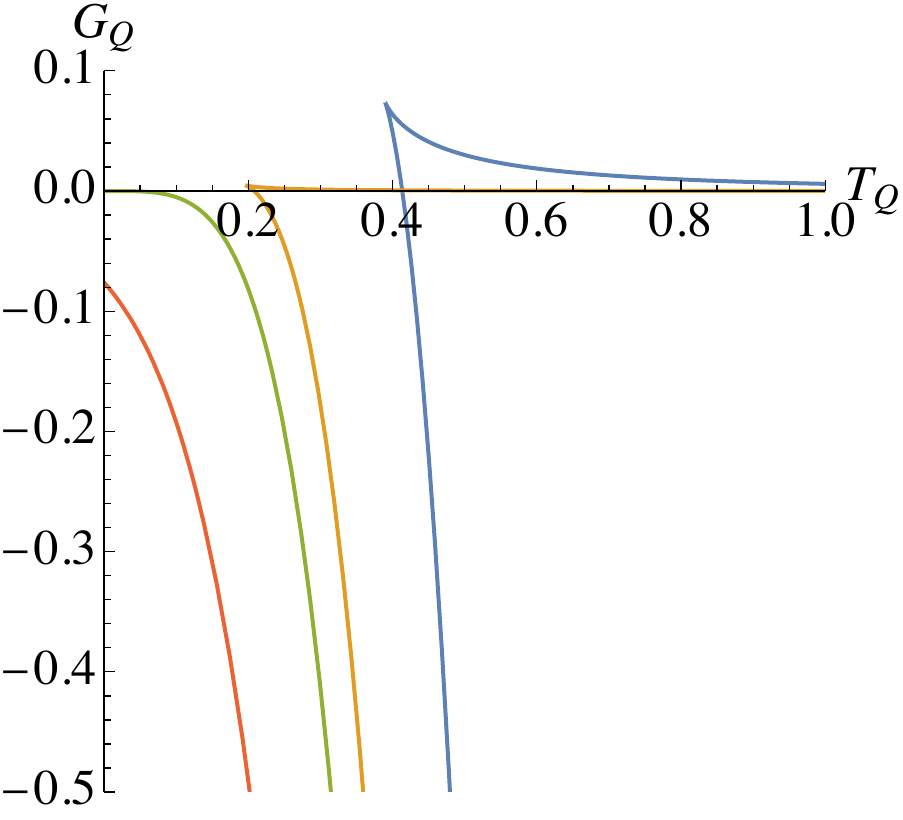}
    \quad
    \includegraphics[scale=0.8]{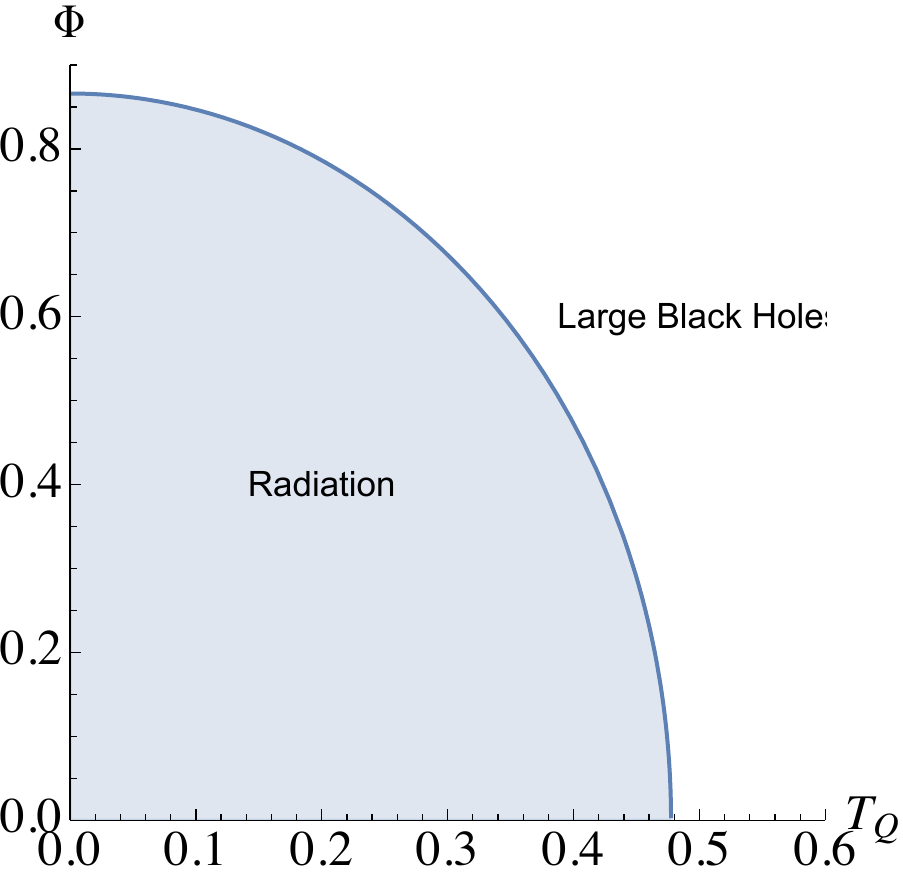}
    \caption{Bulk grand canonical ensemble (at fixed electric potential $\Phi$) of Reissner--Nordstr\"{o}m AdS black holes. \textbf{Left:} Free energy against temperature diagram for $\Phi = 1/2 \Phi_c$ (blue), $\Phi = 9/10 \Phi_c$ (yellow), $\Phi = \Phi_c = \sqrt{3}/2$ (green) and $\Phi = 6/5 \Phi_c$ (red). \textbf{Right:} Co-existence phase  diagram for $\Phi$ vs. $T_Q$.}
    \label{fig:chargedbulk}
\end{figure}

The grand canonical ensemble for the bulk rotating black hole system is similar to that of the boundary CFT. We summarise the main features of the bulk grand canonical ensemble in this Appendix. 

The bulk thermodynamic variables for the   Kerr-AdS black hole can be found in the main text. We restate them here for convenience:
\begin{align}
    M &=  \frac{\Omega_{d-2}}{4 \pi G_N} \frac{m}{\Xi^2} \left ( 1 + \frac{(d-4)\Xi}{2} \right) \,, \quad
     S = \frac{\Omega_{d-2}}{4 G} \frac{r_h^{d-4}(a^2 + r_h^2)}{\Xi} \,,
\\
 	T_H &=  \frac{1}{2\pi} \left [ r_h \left ( 1 + \frac{r_h^2}{L^2}\right)  \left ( \frac{1}{a^2 + r_h^2} + \frac{d-3}{2 r_h^2}\right) -\frac{1}{r_h}\right]\,,
\\
   \Omega &= \frac{a}{L^2} \frac{r_h^2 + L^2}{r_h^2 + a^2} 
   \,, \qquad
   J = \frac{\Omega_{d-2}}{4 \pi G_N} \frac{ma}{\Xi^2} 
\,, \qquad
   P = \frac{(d-1)(d-2)}{16 \pi G_N L^2} \,.
\end{align}
The associated free energy in the grand canonical ensemble is: $W_{\text{bulk}}  = M - T_H S - \Omega J$. As in the main text, we study the phase behaviour of the system by looking at the free energy against temperature diagram, plotted in Fig. \ref{fig:bulkgrand} for $d=4$. As in the CFT case, there is a threshold value, $\Omega L = 1$ (cf.~$\tilde \Omega R = 1$ for the CFT), at which the diagram experiences a qualitative change: for $\Omega L < 1$ the $W_{\text{bulk}} - T_H$ curve displays a Hawking-Page like cusp, while for $\Omega L > 1$ the $W_{\text{bulk}} - T_H$ curve is smooth and positive. For $\Omega L = 1$, the curve is also smooth and positive, but terminates at finite temperature, where $a=L$, beyond which the solution becomes unphysical  with $a>L$ (dashed in figure).

As in the CFT case, the behaviour of the rotating black hole in the grand canonical ensemble is somewhat different from that of the charged black hole, illustrated in Fig. \ref{fig:chargedbulk} (see also e.g. \cite{Chamblin:1999tk}). In particular, the free energy of the charged system also transitions from a cusp to a smooth curve at some (electric) potential $\Phi = \Phi_c=\sqrt{3}/2$,  like in the rotating case. However,   unlike the rotating case,  the free energy curve with $\Phi\geq \Phi_c$ is \textit{negative} and the ``large black hole'' phase   always dominates  over the radiation phase  for these values of the potential in the grand canonical ensemble.

\section{Phase diagram of fixed ($J,\vol,\mu$) ensemble}\label{appB}
In this appendix, we include more details on the $\mu-T$ phase diagrams in Figs. \ref{fig:muT4d} and \ref{fig: GT6DphaseD}. 
As explained in the main text, the phase behaviour of the system is implied by the $G-T$ plots. By studying these in detail,   the various zeroth-order phase transition curves on the $d=4$ phase diagram (Fig.~\ref{fig:muT4d}) can be obtained as follows. 

\begin{itemize}
    \item[1.] $\mu>0$, NS-LE (green): the  smallest real root of $G=0$ (i.e., $x^3-x+ R\mu = 0$), where the free energy $G$ is given by \eqref{G},  is  
    $$x=\frac{2}{\sqrt{3}} \cos \left(\frac{4 \pi}{3}+\frac{1}{3} \cos ^{-1}\left(-\frac{3}{2} \sqrt{3} \mu  R\right)\right)
    \,.$$
   The $T(\mu)$ equation of this zeroth-order phase transition curve can then be obtained directly by substituting this into~\eqref{T3}.
   \item[2.] $\mu>0$, LE-HE and $\mu<0$, NS-HE (yellow): the  second real root of $G=0$ is
   $$x=\frac{\sqrt[3]{2} \left(\sqrt{81 \mu ^2 R^2-12}-9 \mu  R\right)^{2/3}+2 \sqrt[3]{3}}{6^{2/3} \sqrt[3]{\sqrt{81 \mu ^2 R^2-12}-9 \mu 
   R}}\,,$$
   \item[3.] $\mu>0$, NS-HE (orange): the solution to $\partial G/\partial x|_{\mu,R}=0$ is given by $$x=\frac{1}{4} \left(\mu  R \left(\frac{\mu  R}{\sqrt[3]{\mu ^3 R^3+4 \sqrt{\mu ^2 R^2 \left(\mu ^2 R^2+4\right)}+8 \mu 
   R}}+1\right)+\sqrt[3]{\mu ^3 R^3+4 \sqrt{\mu ^2 R^2 \left(\mu ^2 R^2+4\right)}+8 \mu  R}\right)\,.$$
   \item[4.] The intersection of the above three lines     takes place at 
   $$\mu=\frac{2}{3\sqrt{3}R}\,,\qquad x=\frac{1}{\sqrt{3}}\,.$$
   \item[5.] $\mu>0$ HE-IE and $\mu<0$ HE-NS (blue): $z=1$ point, given by $x=1$ at which $T=\frac{1}{2\pi R}$.
   \item[6.] The intersection of 
   the curve in item (3) above with that in item (5)   occurs at
   $$x=1\,, \qquad \mu=\frac{1}{R}\,.$$
\end{itemize}
The
$d=6$ phase diagram (figure \ref{fig: GT6DphaseD}) can be analysed in an analogous way, but in this case the solutions are mostly roots of polynomials of order $\geq5$ for which no analytic expressions exist. Qualitatively:
\begin{itemize}
    \item[1.] $\mu>0$, LE-IE (yellow), $\mu>0$, IE-HE and $\mu<0$, NS-HE (orange): these are different solutions to the equation $G(x)=0$  at given ($\mu,R$).
    \item[2.] $\mu>0$, IE-NS (blue) and $\mu>0$, LE-HE (green): these are different solutions to $\partial G/\partial x|_{\mu,R}=0$.
    \item[3.] The intersection of the LE-IE and IE-HE transition lines happens at $$x=\sqrt{\frac{3}{5}}\,,\qquad \mu = \frac{6}{25R}\sqrt{\frac{3}{5}}\,,$$
    \item[4.] The (purple) vertical line with $z=1$ occurs at $x=1$, giving $T=\frac{3}{2\pi R}$\,.
\end{itemize}

\bibliography{main}

\end{document}